\documentclass[twocolumn,superscriptaddress,nofootinbib,showpacs,preprintnumbers]{revtex4-1}

\usepackage[latin9]{inputenc}
\usepackage{units}
\usepackage{amsmath}
\usepackage{graphicx}
\usepackage{amssymb}
\usepackage{amsfonts}
\usepackage{verbatim}
\usepackage{natbib}
\usepackage[english]{babel}
\usepackage{url}

\usepackage[english]{babel}

\usepackage[usenames,dvipsnames]{color}
\usepackage[caption=false]{subfig} 
\usepackage{upgreek}

\newcommand{\bildb}[7]{
\begin{figure}[#7]
    \center
    \includegraphics[width=#2]{#1}
    \parbox[t]{#4}{\caption[#6]{#3}\label{#5}}
\end{figure}
} 

\def\thepage{\arabic{page}}
\begin{document}

\title{Measurement of $\alpha$--par\-ticle quenching in LAB based scintillator in independent small--scale experiments}

\author{B.~von~Krosigk}
\email{\url{bkrosigk@physics.ubc.ca}}
	\affiliation{Technische Universit\"at Dresden, Institut f\"ur Kern-- und Teilchenphysik, D--01069 Dresden, Germany}
	\affiliation{University of British Columbia, Department of Physics \& Astronomy, Vancouver, BC V6T 1Z1, Canada}
\author{M.~Chen}
        \affiliation{Queen's University, Department of Physics, Engineering Physics \& Astronomy, Kingston, ON K7L 3N6, Canada}
\author{S.~Hans}
        \affiliation{Brookhaven National Laboratory, Upton, NY 11973, USA}
        \affiliation{Bronx Community College, Bronx, NY 10453, USA}
\author{A.R.~Junghans}
        \affiliation{Helmholtz--Zentrum Dresden--Rossendorf, D--01314 Dresden, Germany}
\author{T.~K\"ogler}
	\affiliation{Technische Universit\"at Dresden, Institut f\"ur Kern-- und Teilchenphysik, D--01069 Dresden, Germany}
         \affiliation{Helmholtz--Zentrum Dresden--Rossendorf, D--01314 Dresden, Germany}
\author{C.~Kraus}
        \affiliation{Queen's University, Department of Physics, Engineering Physics \& Astronomy, Kingston, ON K7L 3N6, Canada}
        \affiliation{Laurentian University, 935 Ramsey Lake Road, Sudbury, ON P3E 2C6, Canada}
\author{L.~Kuckert}
	\affiliation{Technische Universit\"at Dresden, Institut f\"ur Kern-- und Teilchenphysik, D--01069 Dresden, Germany}
	\affiliation{Karlsruher Institut f\"ur Technologie, Institut f\"ur Experimentelle Kernphysik, D--76131 Karlsruhe, Germany}
\author{X.~Liu}
        \affiliation{Queen's University, Department of Physics, Engineering Physics \& Astronomy, Kingston, ON K7L 3N6, Canada}
\author{R.~Nolte}
        \affiliation{Physikalisch--Technische Bundesanstalt, Bundesallee 100, D--38116 Braunschweig, Germany}
\author{H.M.~O'Keeffe}
        \affiliation{Queen's University, Department of Physics, Engineering Physics \& Astronomy, Kingston, ON K7L 3N6, Canada}
        \affiliation{Lancaster University, Physics Department, Lancaster, LA1 4YB, United Kingdom}
\author{H.~Wan~Chan~Tseung}
        \affiliation{University of Washington, Center for Experimental Nuclear Physics and Astrophysics, and Department of Physics, Seattle, WA 98195, USA}
        \affiliation{Mayo Clinic, Department of Radiation Oncology, Rochester, MN 55905, USA}
\author{J.R.~Wilson}
        \affiliation{Queen Mary, University of London, School of Physics and Astronomy, London E1 4NS, United Kingdom}
\author{A.~Wright}
        \affiliation{Queen's University, Department of Physics, Engineering Physics \& Astronomy, Kingston, ON K7L 3N6, Canada}
\author{M.~Yeh}
        \affiliation{Brookhaven National Laboratory, Upton, NY 11973, USA}
\author{K.~Zuber}
	\affiliation{Technische Universit\"at Dresden, Institut f\"ur Kern-- und Teilchenphysik, D--01069 Dresden, Germany}


\begin{abstract}
The $\alpha$--par\-ticle light response of liquid scintillators based on linear alkylbenzene (LAB) has been measured with three different experimental approaches. In the first approach, $\alpha$--par\-ticles were produced in the scintillator via $^{12}$C($n$,$\alpha$)$^9$Be reactions.  In the second approach, the scintillator was loaded with 2\% of $^{\mathrm{nat}}$Sm providing an $\alpha$--emitter, $^{147}$Sm, as an internal source.  In the third approach, a scintillator flask was deployed into the water--filled SNO+ detector and the radioactive contaminants $^{222}$Rn, $^{218}$Po and $^{214}$Po provided the $\alpha$--par\-ticle signal. The behavior of the observed $\alpha$--par\-ticle light outputs are in agreement with each case successfully described by Birks' law. The resulting Birks parameter $kB$ ranges from $(0.0066\pm0.0016)$\,cm/MeV to $(0.0076\pm0.0003)$\,cm/MeV. In the first approach, the $\alpha$--par\-ticle light response was measured simultaneously with the light response of recoil protons produced via neutron--proton elastic scattering. This enabled a first time a direct comparison of $kB$ describing the proton and the $\alpha$--par\-ticle response of LAB based scintillator. The observed $kB$ values describing the two light response functions deviate by more than $5\sigma$. The presented results are valuable for all current and future detectors, using LAB based scintillator as target, since they depend on an accurate knowledge of the scintillator response to different particles. 
\end{abstract}

\maketitle

\section{Introduction}
\label{sec:intro}
Over the last decades, liquid scintillation detectors \linebreak gained great importance in neutrino and astroparticle physics. This is due to their capability to detect the charged secondary particles of neutrino interactions down to energies of a few keV in realtime and due to the easy scaling to large target masses. The scintillation light output $L$ scales with the energy of the charged particle, providing valuable kinematic information. In addition, the observed light output of the scintillator at a certain particle energy decreases with increasing ionization density. This effect, known as ionization quenching, allows to discriminate heavily ionizing particles, like protons and $\alpha$--particles, from electrons. The energy dependent, quenched scintillation light output can be described analytically by Birks law \cite{bir51, bir64},
\begin{equation}
\label{equ:birks}
L(E) = S \cdot \int_{0}^{E} \frac{\mathrm{d}E}{1+kB\left(\frac{\mathrm{d}E}{\mathrm{d}x}\right)},
\vspace{1mm}
\end{equation}
where  $\mathrm{d}E/\mathrm{d}x$ is the specific energy loss. $kB$ denotes Birks' parameter and $S$ is a scaling parameter, which is associated with the scintillation efficiency. For fast electrons, which have a small $\mathrm{d}E/\mathrm{d}x$, Eq.~\ref{equ:birks} approximates the proportionality
\begin{equation}
\label{equ:eresp}
L_e(E) = S\cdot E,
\end{equation}
where the index $e$ refers to electrons \cite{bir64}. For many of the standard organic liquid and plastic scintillators, a linear electron scintillation light response has been observed down to about 100\,keV \cite{bra62,abe11,nov97,wan11,koe13}. This includes also LAB based scintillators \cite{wan11,koe13}. At lower energies, $\mathrm{d}E/\mathrm{d}x$ is increased with respect to fast electrons and $L$ rises non--linearly with $E$. The small non--linearity results in an energy offset, if Eq.~\ref{equ:eresp} is used to describe the electron light output function. This offset was experimentally determined to be $\lesssim5$\,keV \cite{abe11,nov97a,wan11}. This quasi--linear be\-ha\-vior of scintillation light induced by electrons is ty\-pi\-cal\-ly taken advantage of in order to calibrate the light output scale in ionization quenching measurements. \mbox{Often} a scale in units of electron--equivalent energy is chosen and thus $S=1$. This scale is used throughout this article for $L$.

With the evolution of both, detector technology and understanding of the scintillator properties, experiments using scintillation detectors grew in the second half of the last century from small--scale to multi--tonne detectors. Several large--scale liquid scintillator detectors with up to 1\,kt scintillator mass are currently operational worldwide and multi--kilotonne devices are \mbox{being} designed. However, despite the great progress in understanding the mechanisms of liquid scintillation, some fundamental questions still remain unanswered. One of these questions is whether the same value of $kB$ can describe the light response of a certain organic scintillator to dif\-fe\-rent ions \cite{tre13}, a question that was raised already in the original works by J.B.~Birks \cite{bir64} upon the observation of $kB(\mathrm{proton})\neq kB(\alpha)$. However, since no consistent behavior was observed, no conclusive answer was found at that time. And also to date, this question still remains unresolved. As already discussed in the seminal work of Birks \cite{bir51, bir64}, the increased complexity of the reaction kinetics in multi--component \mbox{liquid} scintillators compared with organic plastic scintillators or inorganic scintillators makes a unique \mbox{answer} to this question impossible. Therefore, individual investigations are crucial for the wide range of operational and future large--scale liquid scintillator detectors like Daya Bay \cite{an12}, RENO \cite{ahn12}, SNO+ \cite{loz15}, JUNO \cite{zha13}, RENO--50 \cite{kim14}, LENS \cite{kor01} and HANOHANO \cite{cic12}. All of these experiments use, or consider to use, LAB based scintillator.

Section~\ref{sec:nbeam} of this article describes the measurement and analysis of $\alpha$--par\-ticle quenching in LAB based scintillators using fast neutrons. This experiment is referred to as the "neutron beam experiment". The measurements were carried out at the PTB\footnote{Physikalisch--Technische Bundesanstalt.} Ion Accelerator Facility (PIAF) \cite{bre80,kle80}, which provides a neutron beam with a continuous energy distribution. The individual neutron energies are derived from time--of--flight (TOF) measurements. $\alpha$--par\-ticles are produced inside the scintillator through ($n$,$\alpha$) reactions and outgoing $\alpha$--parti\-cles with an energy known from kinematic calculations are identified within the data. The obtained quen\-ching results are compared to the results of the proton quen\-ching measurements published in \cite{kro13}, which make use of $n$--$p$ elastic scattering in the scintillator. Both measurements were taken simultaneously and with the same scintillation detector. 

Section~\ref{sec:SmLAB} presents the second $\alpha$--quen\-ching experiment, which uses sa\-ma\-rium--loaded LAB based scintillator, referred to as the "samarium experiment". The scintillator was loaded for this experiment with 2\% \linebreak $^{\mathrm{nat}}$Sm at BNL\footnote{Brookhaven National Laboratory.} and the measurement was carried out at HZDR\footnote{Helmholtz--Zentrum Dresden--Rossendorf.}. The isotope $^{147}$Sm is an $\alpha$--emitter with a $Q$--value of 2.3105(11)\,MeV \cite{eks04}, providing an internal $\alpha$--source.

Section~\ref{sec:bucket} describes the third measurement of $\alpha$--par\-ticle quenching in LAB based scintillator, referred to as the "bucket source experiment". In this case, the results were obtained from a 1\,l sample deployed within an acrylic container, the bucket, into the water--filled SNO+ detector. Contaminations of the scintillator by the $\alpha$--emitters $^{222}$Rn, $^{218}$Po and $^{214}$Po are used as internal $\alpha$--sources with particle energies of 5.49\,MeV, 6.00\,MeV and 7.69\,MeV, respectively. 

All three independent experiments were conducted with small liquid scintillator volumes, using charged particles produced in the scintillator itself. Furthermore all three experiments had a comparable sensitivity to UV light. This is important because a small fraction of Che\-ren\-kov light is always emitted in addition to the scintillation light, when an electron with an energy above the Cherenkov threshold of about 166\,keV traverses the LAB based scintillator. In references \cite{wan11,koe13} it is shown that the additional Cherenkov light slightly changes the gradient of the linear relation between electron energy and light output and thus the calibration to electron--equivalent energy. Throughout this article, the observed pulse--height $PH$ is calibrated in units of electron--equivalent energies, including the scintillation and Cherenkov photons.

The scintillator consists of the solvent LAB, the primary fluor 2,5--diphenyloxazole (PPO) and the secon\-dary fluor, if any, p--bis--(o--methylstyryl)--benzene \linebreak (bis--MSB). Bis--MSB acts as a wavelength shifter to achieve a better match of the spectral distribution of the scintillation light and the sensitivity of the photocathode. The LAB solvent was obtained from Petresa Canada Inc. \cite{p500q} with an average stoichiometric composition of C$_{17.1}$H$_{28.3}$. 

A direct comparison of the $\alpha$--par\-ticle quenching measurement results from these three experiments is presented in Sec.~\ref{sec:ThreeAlpha}. Section~\ref{sec:AlphaProton} provides a direct comparison of the quenching parameters $kB$ for protons and $\alpha$--par\-ticles, determined in a simultaneous measurement using neutron--induced reactions in the scintillator. The proton quenching data were already published earlier \cite{kro13}. A summary of the article and an outlook is given in Sec.~\ref{sec:summary}.

\section{Measurements of $\alpha$--par\-ticle quenching using fast neutrons}
\label{sec:nbeam}

In the neutron beam experiment, the $\alpha$--par\-ticle quen\-ching in two deoxygenated, ternary LAB based scintillators is determined from the same data sets used for the proton quenching analysis \cite{kro13}. The respective scintillator samples are LAB + 2\,g/l PPO + 15\,mg/l bis--MSB and LAB + 3\,g/l PPO + 15\,mg/l bis--MSB. The only difference in these two samples is the PPO concentration. The analysis of the $\alpha$--par\-ticle quen\-ching has higher demands on the $PH$ resolution than the one of the proton quenching because structures caused by $\alpha$--par\-ticles from ($n$,$\alpha$) reactions have to be discriminated from those resulting from other reactions. Thus the additional two binary LAB based scintillators without bis--MSB, used in \cite{kro13}, cannot be used for the ana\-lysis of the $\alpha$--par\-ticle quenching because they show a lower resolution than the ternary scintillators. The experimental setup, data acquisition and calibration is described in detail in \cite{kro13}. Therefore only details spe\-ci\-fi\-cally relevant for the analysis of the $\alpha$--par\-ticle quen\-ching are discussed here.

\subsection{Experimental setup and data extraction}
The beam of fast neutrons was produced by bom\-bar\-ding a 3\,mm Beryllium target with 19\,MeV protons from the CV28 cyclotron at PIAF. The neutrons re\-sul\-ting from $^9$Be($p$,$nx$) reactions have a continuous kinetic energy distribution from about 1\,MeV to 17\,MeV and are observed together with prompt $\gamma$--rays produced in inelastic interactions. The individual neutron energy $E_n$ is deduced from a measurement of the neutron TOF relative to the centroid TOF value of the prompt $\gamma$--peak together with the flight distance from the target to the scintillator volume. 

The scintillation detector had an active volume of about 100\,ml. It was observed by an XP2020Q PMT, which has an increased UV light sensitivity. All materials between scintillator and photocathode were UV transparent. An integrated and amplified charge signal was derived from the ninth dynode out of twelve, avoiding a non--linear PMT gain and its pulse height $PH$ was measured using a peak sensing analog--to--digital converter (ADC).

The high--voltage applied to the PMT, and thus the PMT gain, was increased for the $\alpha$--par\-ticle quenching and low energy proton quenching measurements, compared to the proton quenching measurements at energies above about 5\,MeV \cite{kro13}. This achieves a better re\-so\-lu\-tion of low $PH$ signals, which is especially important for the analysis of the highly quenched $\alpha$--par\-ticle light pulses. This mode is referred to as high gain (HG) mode.

The data are stored in a $PH$ versus TOF matrix. From this matrix, $PH$ spectra are extracted by se\-lec\-ting a small TOF window around the TOF of interest and by projection of the selected events on the $PH$ axis.  The $PH$ scale is calibrated with $\gamma$--ray sources yielding $L$ in units of electron--equivalent energy. The TOF, and thus $E_n$, window is always made smaller than the $E_n$ window corresponding to the light output resolution $\Delta L$ at the respective particle energy.

\subsection{Calibration}
\label{ssec:nbeamgcal}

Three $\gamma$--ray sources were used for calibration: $^{137}$Cs, $^{22}$Na and $^{207}$Bi. They provide in total six $\gamma$--rays with different energies. Figure~\ref{fig:Bi207} shows the observed $PH$ distribution of $^{207}$Bi as an example. The three $^{207}$Bi $\gamma$--rays lead to three sharp edges at which the Compton electrons have the maximum energy $E$. Since this energy is precisely known, the relation between $PH$ and electron energy is determined by fitting the simulated $PH$ distributions in the region around the Compton edges to the measured ones \cite{kro13}. The simulation is performed with the code \textsc{gresp7} \cite{die82}. Despite small deviations at lower pulse--heights, the region around the Compton edges is very well described by the simulation. The deviations result mainly from a simplified description of surface effects within the \textsc{gresp7} simulation.

\bildb{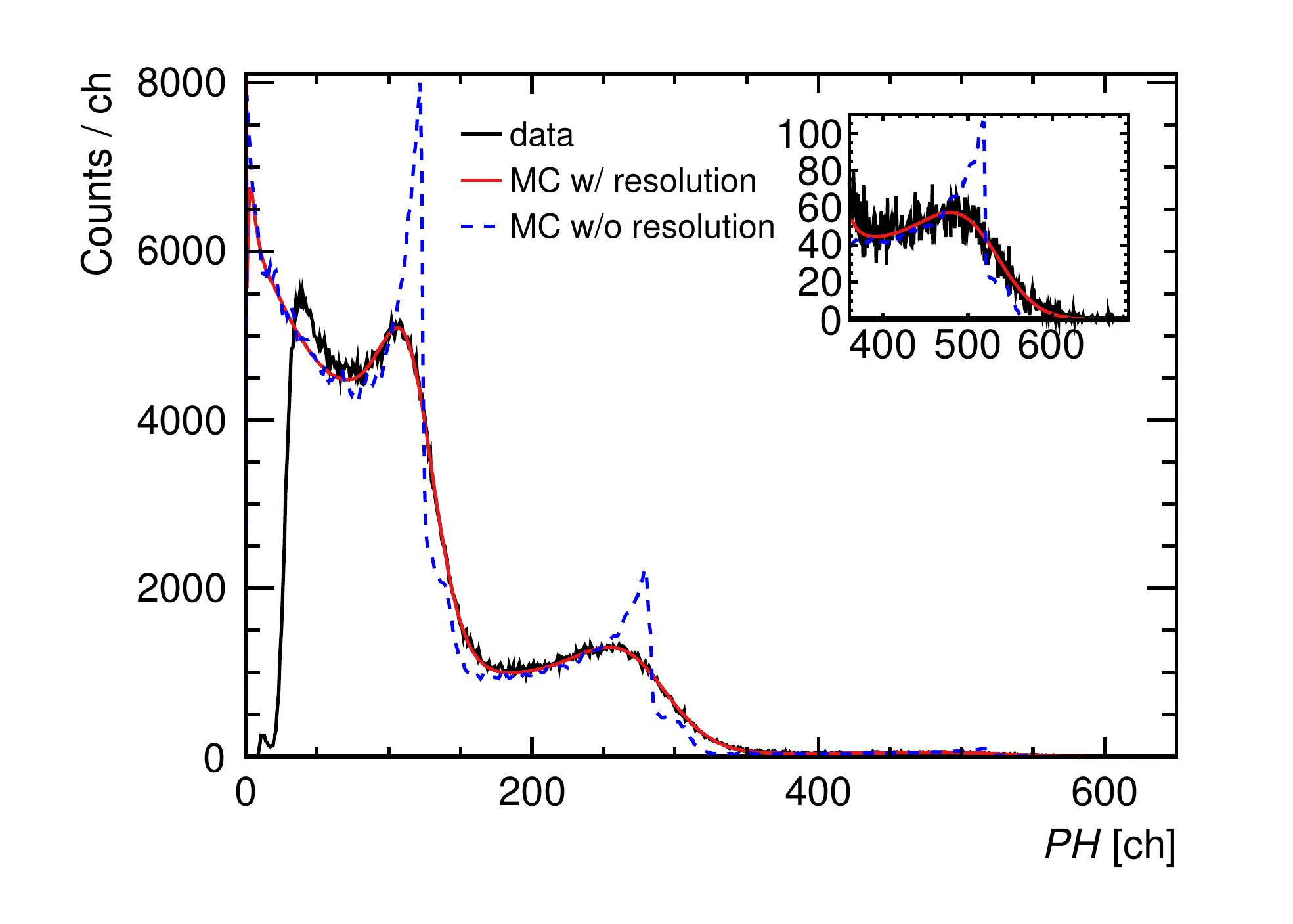}{0.5\textwidth}{Distribution of the pulse--height $PH$ measured in ADC channels with a $^{207}$Bi calibration source. The data were taken with LAB + 2\,g/l PPO + 15\,mg/l bis--MSB. Also shown is the Monte Carlo (MC) distribution before and after folding it with the detector resolution function. The third Compton edge at high $PH$ is shown enlarged in the inset}{0.48\textwidth}{fig:Bi207}{}{h!}

The described procedure yields six data points \linebreak $PH_i = PH(E_i)$, where $i=1-6$. A linear fit to these data points, with 
\begin{equation}
\label{equ:gcal}
PH(E) = m \cdot E + a,
\end{equation}
results in the two calibration parameters $m=(343.4\pm1.1)$\,channels\,/\,MeV and $a=(-9.0\pm2.0)$\,channels for LAB + 2\,g/l PPO + 15\,mg/l bis--MSB. Using instead LAB + 3\,g/l PPO + 15\,mg/l bis--MSB results in the \linebreak parameter values $m=(350.5\pm0.7)$\,channels\,/\,MeV and \linebreak $a=(-11.1\pm2.1)$\,channels. 

The light output in units of electron--equivalent energy is finally obtained from equations~\ref{equ:eresp} and \ref{equ:gcal} and thus
\begin{equation}
\label{equ:Lee}
L(E) = \frac{PH - a}{m}.
\end{equation}

\subsection{Determination of the $\alpha$--par\-ticle response}
\label{ssec:nbeamAlpha}

\subsubsection{Reactions producing $\alpha$--par\-ticles}

The beam neutrons reaching the sensitive volume mostly elastically scatter off protons in the scintillator pro\-du\-cing a recoil proton energy distribution \cite{kro13}. Neutrons that exceed a threshold energy of 6.19\,MeV can furthermore produce $\alpha$--par\-ticles via the reaction 
\begin{equation}
\label{equ:nalpha}
^{12}\mathrm{C}(n,\alpha)\,^9\mathrm{Be}(\mathrm{g.s.})
\end{equation}
with scintillator intrinsic $^{12}$C nuclei. The maximum possible energy of the single $\alpha$--par\-ticle in the final state is unambiguously related to the energy $E_n$ of the incoming neutron. Above $E_n =8.81$\,MeV and 8.29\,MeV, respectively, two further reactions contribute to the production of $\alpha$--par\-ticles:
\begin{align}
\label{equ:2alpha}
^{12}\mathrm{C}(n,\alpha^\prime)\,^9\mathrm{Be}^* \rightarrow\> &n+\,^8\mathrm{Be}\\\nonumber
&\>\>\>\>\>\>\>\>\>\>\> \downarrow\\\nonumber
&\>\>\>\>\>\>\>\>\>\>\>2\alpha ,\\\nonumber
&\\
\label{equ:3alpha}
^{12}\mathrm{C}(n,n')\,^{12}\mathrm{C}^* \rightarrow\> &\alpha +\,^8\mathrm{Be}\\\nonumber
&\>\>\>\>\>\>\>\>\>\>\> \downarrow\\\nonumber
&\>\>\>\>\>\>\>\>\>\>\>2\alpha.
\end{align}
These reactions have more than one $\alpha$--par\-ticle in the final state and it is not possible to identify $\alpha$--par\-ticles of known energy. They thus contribute to the background spectrum.

Natural background, for instance from ambient $\gamma$--ra\-di\-a\-tion and scintillator internal impurities, was measured in the absence of calibration sour\-ces and beam neutrons and subtracted from data.

\subsubsection{Monte Carlo simulations}

The simulation of light yield distributions resulting from incoming neutrons is performed with the Monte Carlo (MC) code  \textsc{nresp7} \cite{nresp}. The non--linearly rising light output from different charged particles is simulated using a set of predefined light output functions, which are stored in an external file and iteratively adapted to the data. \textsc{nresp7} describes the $\alpha$--particle light output function by two analytic expressions:

\begin{align} 
L_{\alpha}(E) &= c_0 E^{c_1} \>\>\>	&E < 6.76\mathrm{\,MeV}, \label{equ:La1_nresp}\\ 
L_{\alpha}(E) &= c_2 + c_3E\>\>\>	&E\geq6.76\mathrm{\,MeV}. \label{equ:La2_nresp}
\end{align}
The parameter values resulting from the last iteration are $c_0=0.030$, $c_1=1.640$, $c_2=-0.518$, $c_3=0.179$ for LAB + 2\,g/l~PPO + 15\,mg/l bis--MSB and $c_0=0.031$, $c_1=1.689$, $c_2=-0.505$, $c_3=0.190$ for LAB + 3\,g/l~PPO + 15\,mg/l bis--MSB. The respective functions are shown in Fig.~\ref{fig:nresp}.
\bildb{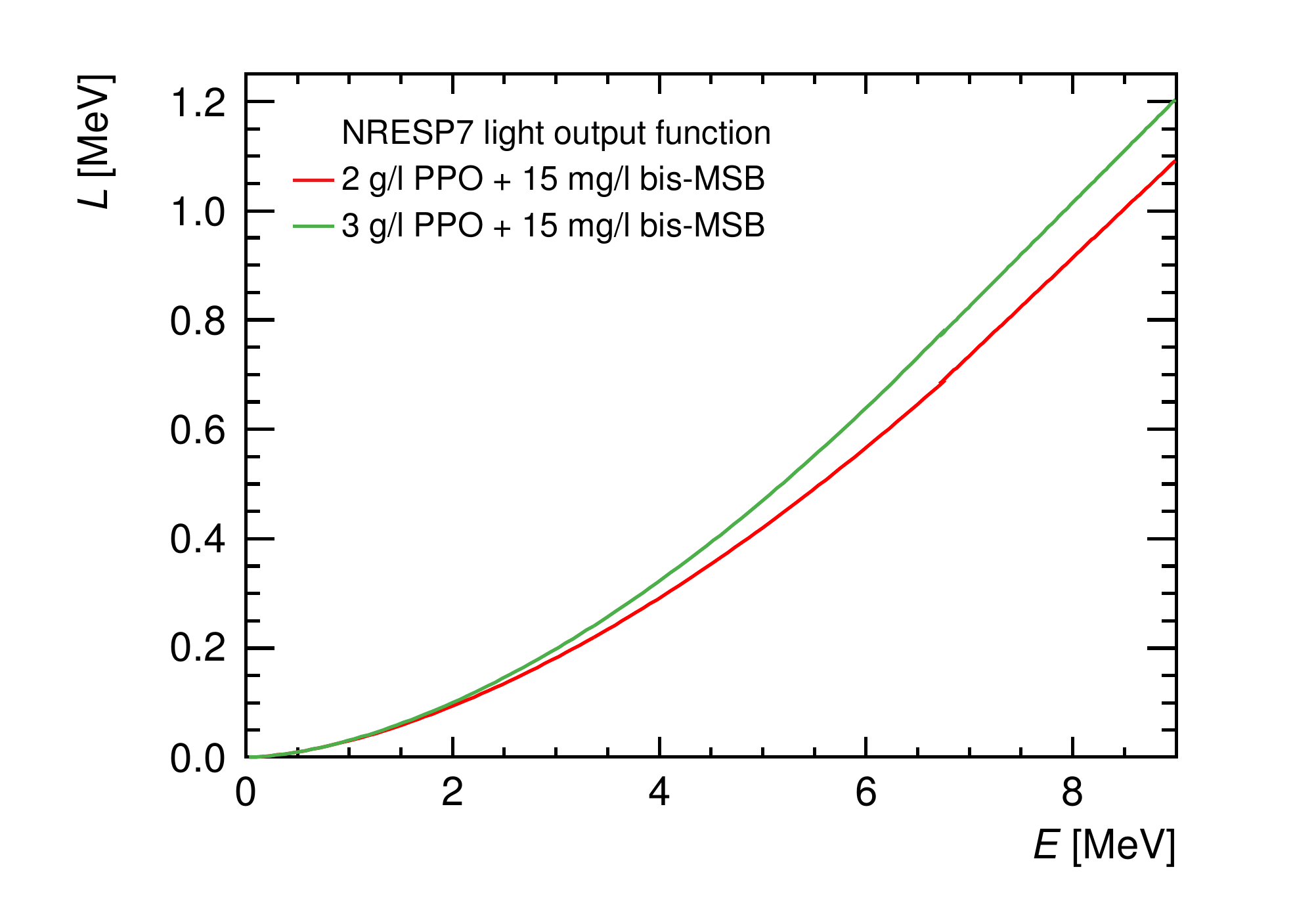}{0.5\textwidth}{Light output $L$ as function of kinetic energy $E$ used in \textsc{nresp7} to simulate $\alpha$--par\-ticle events in LAB + 2\,g/l~PPO + 15\,mg/l bis--MSB and LAB + 3\,g/l~PPO + 15\,mg/l bis--MSB. The two functions shown result from iterative adaptions of the simulated to the measured light yield distributions.}{0.48\textwidth}{fig:nresp}{}{h!}

The determination of the proton light output function is possible without the knowledge of the specific $\alpha$--par\-ticle light output function, which is not the case vice versa. The reason is that the individual regions of interest used for the proton analysis nearly exclusively contain proton events \cite{kro13}. Instead, the regions of interest for the $\alpha$--par\-ticle analysis are not do\-mi\-na\-ted by $\alpha$--par\-ticle, but also by proton events. Hence, the proton analysis was conducted first. 

\bildb{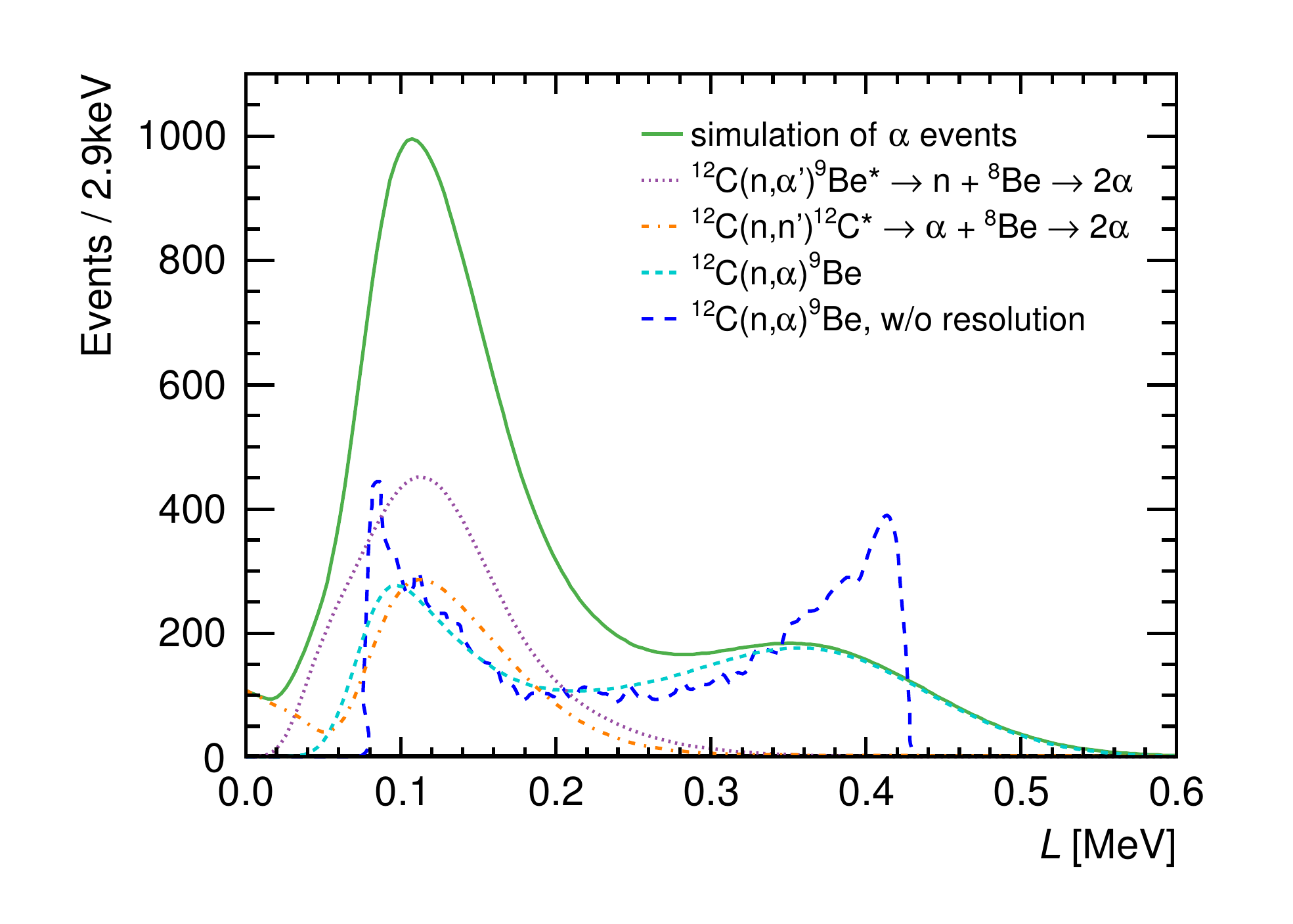}{0.5\textwidth}{Simulated distributions of the total light yield $L$ in units of electron--equivalent energy in LAB + 2\,g/l~PPO + 15\,mg/l bis--MSB due to $\alpha$--par\-ticles from reactions (\ref{equ:nalpha})--(\ref{equ:3alpha}). The energy of the incoming neutrons ranges from 10.6\,MeV to 11.4\,MeV with a uniform energy distribution. Additionally shown is the $L$ distribution due to reaction (\ref{equ:nalpha}), induced by mono--energetic 11\,MeV neutrons and without considering the detector resolution. The sharp high energy edge at about 430\,keV is produced by $\alpha$--par\-ticles with $E_{\alpha}=5.14$\,MeV}{0.48\textwidth}{fig:aspec_1}{}{h!}

Figure~\ref{fig:aspec_1} shows the simulated light yield distributions resulting from the reactions (\ref{equ:nalpha})--(\ref{equ:3alpha}) as well as their sum. The calibration parameters for LAB + 2\,g/l~PPO + 15\,mg/l bis--MSB (see Sec.~\ref{ssec:nbeamgcal}) are considered and an incoming neutron energy of $E_n = (11.0\pm0.4)$\,MeV is assumed. The width of the rectangular neutron energy distribution corresponds to the TOF window width used for the extraction of the respective data spectrum. The edge visible around 430\,keV results from $\alpha$--par\-ticles which are produced in the reaction (\ref{equ:nalpha}) 
and emitted at forward angles. The corresponding $\alpha$--par\-ticle energy can be calculated using simple two--particle kinematics, assuming $^{12}$C being at rest. For the given example with $E_n = 11.0$\,MeV, the $\alpha$--par\-ticle energy is 5.14\,MeV. In Fig.~\ref{fig:aspec_1} the high energy edge lies outside the region \mbox{covered} by events from the break--up reactions (\ref{equ:2alpha}) and (\ref{equ:3alpha}). This enables a precise assignment of a known $\alpha$--par\-ticle energy to a particular value of $L$.

The advantage of \textsc{nresp7} is the relatively detailed description of the ($n$,$\alpha x$) reactions on carbon nuclei, compared to e.g. standard Geant4. There is no Geant4 version available, or dedicated class for the use with Geant4, which contains for instance a model for the various 3$\alpha$--break--up channels\footnote{Work is underway at CIEMAT, Madrid, to migrate the \textsc{nresp7} model to Geant4.}. The ($n$,$\alpha x$) reactions are of par\-ti\-cu\-lar relevance for the determination of the light output function for $\alpha$--par\-ticles. It should be noted, however, that independent of the code, the individual differential neutron cross sections on carbon are still not well--known \cite{nresp}. The available evaluated data files, such as ENDF, JENDL or JEFF, only contain differential level cross sections for $^{12}$C($n$,$\alpha$)$^9$Be$^*$ or double--differential $\alpha$--emission cross sections $\mathrm{d}\sigma/(\mathrm{d}\Omega \mathrm{d}E)$, i.e. the statistical average of the $\alpha$--particle emission over the individual 3$\alpha$--break--up reactions. For modeling the scintillation detector, however, the correlations of the $\alpha$--particles in an individual reaction are required, i.e. $\mathrm{d}\sigma/(\mathrm{d}E_{\alpha 1} \mathrm{d}E_{\alpha 2} \mathrm{d}E_{\alpha 3})$. Although the ENDF format basically has the capability to represent these data, no information on these channels is available for $^{12}$C. The shortcoming regarding the individual differential neutron cross sections on carbon results in an uncertainty on the shape of the $\alpha$--par\-ticle energy distributions. This uncertainty has to be considered for the determination of the position of the $\alpha$--par\-ticle edge from reaction (\ref{equ:nalpha}) within the total light yield distribution.

\subsubsection{Comparison with data}
\label{ssec:beamdata}

The simulated total light yield distribution, re\-sul\-ting from $\alpha$--par\-ticle events and non--$\alpha$--par\-ticle events, is shown in Fig.~\ref{fig:aspec_2} together with the data spectrum. The $\alpha$--par\-ticle edge in the given example is well reproduced by the simulation. The small mismatch between simulation and data, observed in Fig.~\ref{fig:aspec_2}, is due to the fact that \textsc{nresp7} does not model interactions of de--excitation $\gamma$--rays, which result mostly from the first excited state in $^{12}$C at 4.439\,MeV, populated by inelastic neutron scattering. Instead, it is assumed that these events are suppressed by pulse--shape discrimination (PSD). Because of the poor $n/\gamma$--PSD properties of LAB based scintillators, however, $\gamma$--induced events could not be separated from neutron--induced events in the data \cite{kro13}. Hence, a direct comparison of experimental and simulated light yield distributions is difficult in regions where $\gamma$--induced events contribute significantly. Fortunately, the Compton edge from this $\gamma$--radiation lies far outside the region of the observed edge attributed to $\alpha$--par\-ticles. This is obvious from a comparison with the inset in Fig.~\ref{fig:Bi207}, which shows the edge position from $\gamma$--rays with an energy of 1.770\,MeV. Within the $\alpha$--par\-ticle edge region itself, no data excess is visible since the $\alpha$--par\-ticle event simulation is normalized individually to data in this region to reduce the sensitivity to the insufficient knowledge of the neutron cross sections on carbon. As a consequence, the $\alpha$--par\-ticle event yield is slightly overestimated in the presence of background events. The location of the edge, though, is not affected, as long as the background is smooth as is the case discussed here.
\bildb{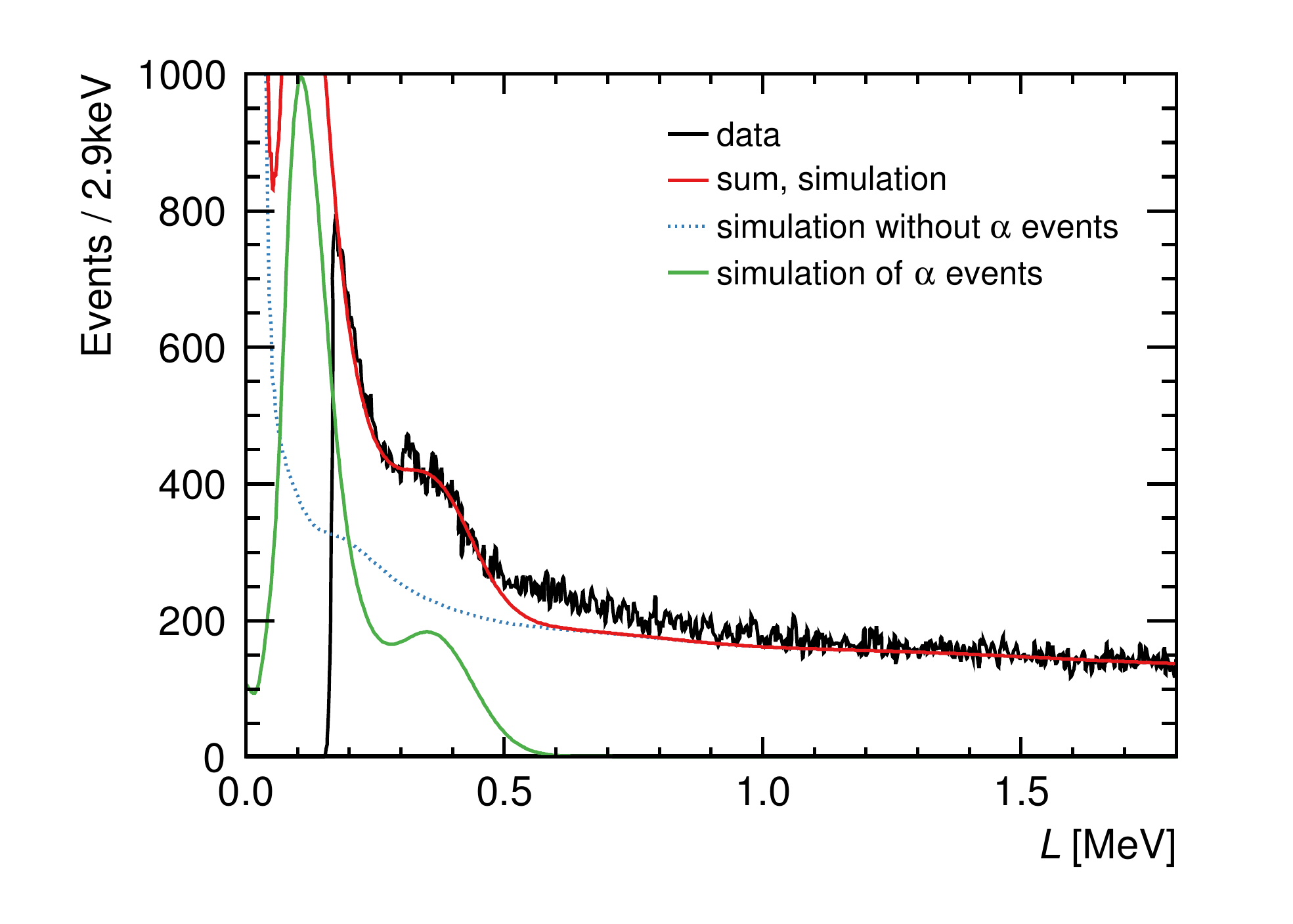}{0.5\textwidth}{Measured and simulated distributions of the total light yield $L$ in units of electron--equivalent energy in LAB + 2\,g/l~PPO + 15\,mg/l bis--MSB induced by neutrons with \mbox{$E_n = (11.0\pm0.4)$\,MeV}. The contributions of $\alpha$--par\-ticle and non--$\alpha$--par\-ticle (mainly recoil proton) events to the total simulated $L$ spectrum are shown individually}{0.48\textwidth}{fig:aspec_2}{}{h!}

The analysis of multiple, measured light yield distributions induced by neutrons with different $E_n$ provides a set of data points $L_i$. This is, of course, only possible if the $\alpha$--par\-ticle edge is identified within the total light yield distribution, which is dominated by recoil proton events. The $^{12}$C($n$,$\alpha$)$^9$Be reaction leads to an observable structure in the $L$ spectra (see Fig.~\ref{fig:aspec_2}), as soon as the $\alpha$--par\-ticle light output is high enough to overcome the detector threshold. This is the case for $E_n \gtrsim 9.5$\,MeV in the investigated data sets. At $E_n \gtrsim 11.5$\,MeV the two additional reactions, (\ref{equ:2alpha}) and (\ref{equ:3alpha}), dominate the $PH$ spectrum resulting from all $\alpha$--par\-ticle events. The $\alpha$--par\-ticle edge with known energy can not be located anymore until $E_n \approx 14$\,MeV. At these high neutron energies, the maximum $\alpha$--par\-ticle energy resulting from reaction (\ref{equ:nalpha}) is high enough to induce a light output well outside the light yield distributions from the other two reactions. Finally, the number of light yield distributions with an observable $\alpha$--par\-ticle edge also depends on the $PH$ re\-so\-lu\-tion. 

The sources of systematic uncertainties are the $PH$ ca\-li\-bra\-tion, the TOF measurement, the PMT gain and the MC simulations. The uncertainties on the calibration parameters $m$ and $a$ are given in Sec.~\ref{ssec:nbeamgcal}. These uncertainties, as well as all uncertainties on the TOF measurement and the PMT gain, are the same as for the measurement of the light production by protons and are discussed in \cite{kro13}. The MC uncertainty, however, differs and is sig\-ni\-fi\-cant\-ly larger for $\alpha$--par\-ticles. It results mainly from the uncertainty on the differential ($n$,$\alpha x$) cross sections for carbon, the effect of photon induced events not \mbox{covered} by the simulation and the residual uncertainty on the determined proton light output function. These uncertainties within the MC manifest in the dependence of the $\alpha$--par\-ticle edge position on the $L$ interval chosen for the fit of the MC to data. For this reason, the $L$ interval around the edge position is systematically varied and the average edge position is determined. The maximum deviation from the average position observed in all analyzed light yield distributions is $\pm26$\,keV.

\subsubsection{Results}
\label{ssec:nbeamresults}

\begin{figure}
	\centering
	\subfloat[LAB\,+\,2\,g/l\,PPO\,+\,15\,mg/l\,bis--MSB]{\label{fig:Ba_LAB:1}\includegraphics[width=0.5\textwidth]{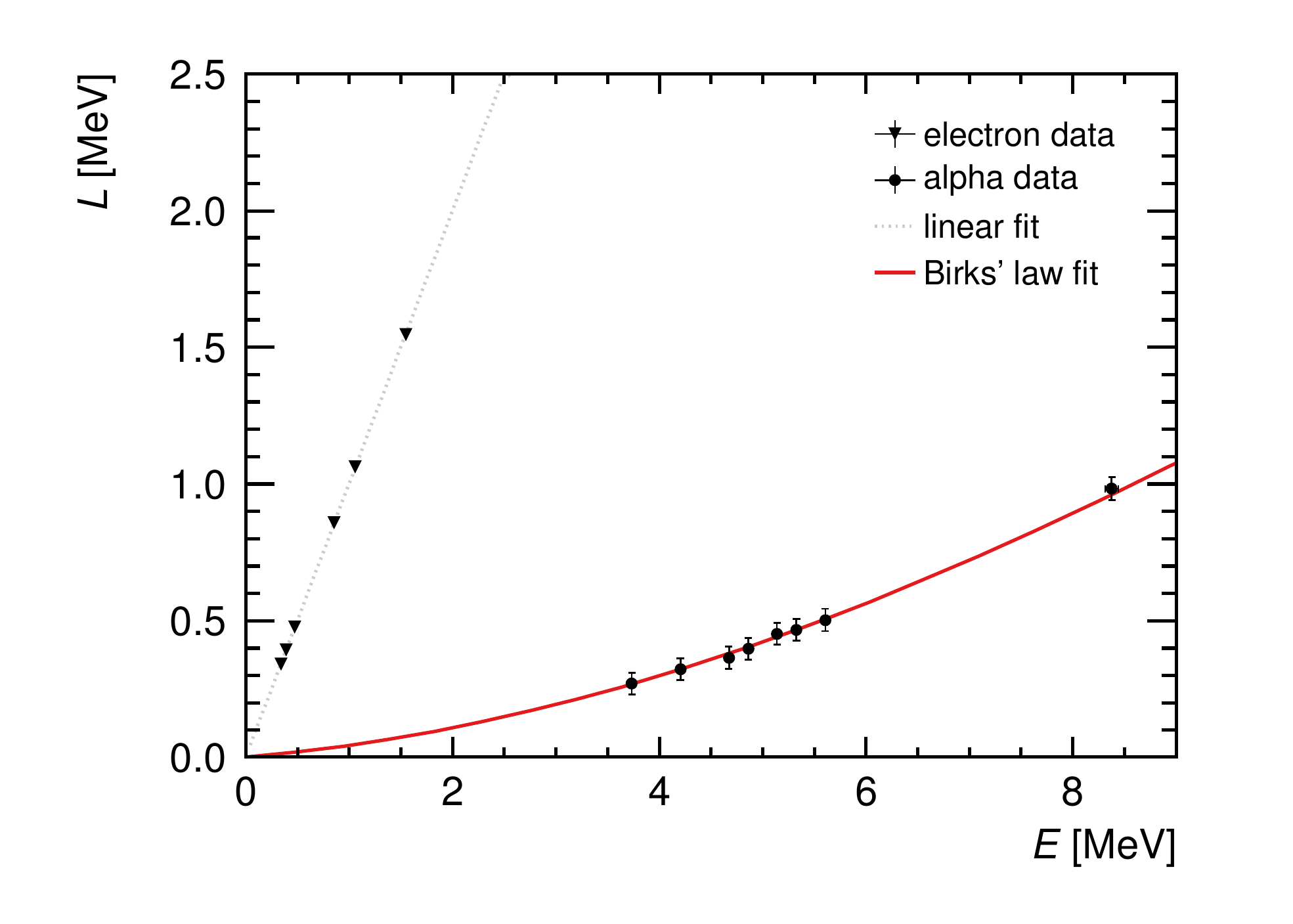}}\qquad
	\subfloat[LAB\,+\,3\,g/l\,PPO\,+\,15\,mg/l\,bis--MSB]{\label{fig:Ba_LAB:2}\includegraphics[width=0.5\textwidth]{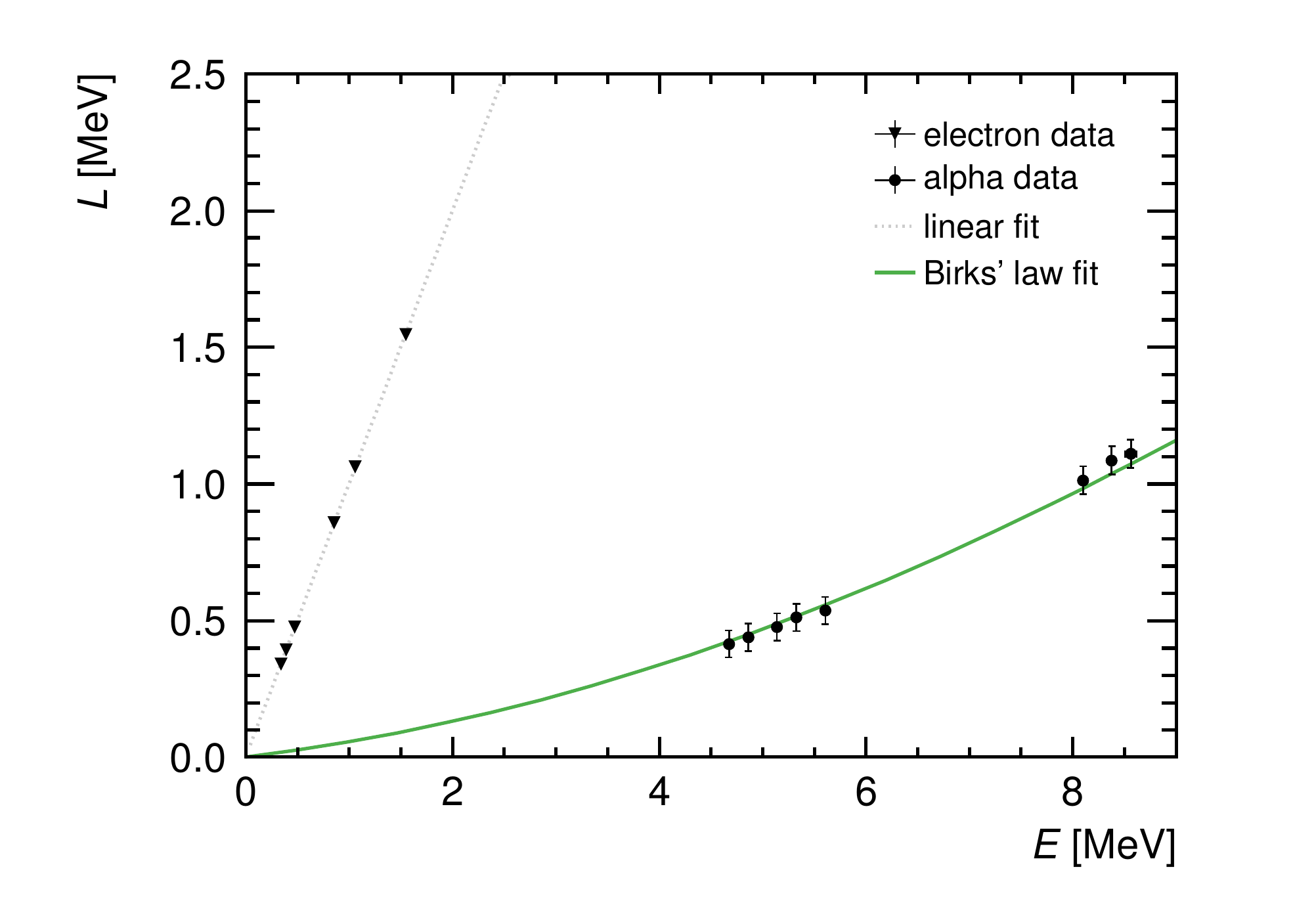}}
	\caption{\label{fig:Ba_LAB} Light output $L$ in units of electron--equivalent energy as function of kinetic energy $E$. $L$ is shown for electrons and $\alpha$--par\-ticles in LAB scintillator with 15\,mg/l bis--MSB and 2\,g/l (a) or 3\,g/l PPO (b). The $\alpha$--par\-ticle data is fitted with Eq.~\ref{equ:birks} including systematic effects as nuisance parameters. In the shown total uncertainties, the single contributions are added quadratically}
\end{figure}

Figures~\ref{fig:Ba_LAB:1} and \ref{fig:Ba_LAB:2} show the $\alpha$--par\-ticle data points $L_i$, determined as described above for LAB + 2\,g/l PPO + 15\,mg/l bis--MSB and LAB + 3\,g/l PPO + 15\,mg/l bis--MSB, respectively. The evaluated $\alpha$--par\-ticle energies partly differ for the two scintillator samples, since besides the fluor concentration, the experimental settings like the amplification were slightly different in the two measurements. Both aspects affect the light output re\-so\-lu\-tion.

The $\alpha$--par\-ticle light output function $L(E)$ is determined by a $\chi^2$ fit of Eq.~\ref{equ:birks} to the data points. In this fit, all systematic uncertainties mentioned in Sec.~\ref{ssec:beamdata} are included as nuisance parameters (see \cite{kro13} for details). The specific energy loss in the respective scintillator is calculated using \textsc{srim} \cite{srim}. An uncertainty on the $(\mathrm{d}E/\mathrm{d}x)(E)$ calculation of 2\% is considered \cite{kro13,zie99}. The resulting best fit values for the Birks parameter are \linebreak \mbox{$kB = (0.0076 \pm 0.0003)$\,cm/MeV} (see Fig.~\ref{fig:Ba_LAB:1}) and \linebreak $kB = (0.0071 \pm 0.0003)$\,cm/MeV (see Fig.~\ref{fig:Ba_LAB:2}). The corresponding $\chi^2$ values per degree of freedom are 0.73 and 1.36, respectively.

These results, obtained using the two LAB samples which only differ in the concentration of the primary fluor, agree within $1.2\,\sigma$. This is in line with the theo\-re\-ti\-cal expectation \cite{bir64} as well as the observation in \cite{kro13} that the fluors do not significantly affect the ionization quenching parameters at the given concentrations. Ioni\-za\-tion quenching is, according to \cite{bir64}, a primary process and thus competing with the excitation of the solvent and not with the energy transfers from the solvent to the solute, which are referred to as secondary processes.

\section{Measurement of $\alpha$--par\-ticle quenching using samarium--loaded scintillator}
\label{sec:SmLAB}
While the $\alpha$--par\-ticle kinetic energy corresponding to an observed light output is laborious to access in the neutron beam experiment presented in Sec.~\ref{sec:nbeam}, its determination is straight forward observing the peak from an $\alpha$--source with known $\alpha$--par\-ticle energy. This requires the use of an internal $\alpha$--source to avoid energy losses within the source carrier and the escape of scintillation light produced near the scintillator surface. For this reason, LAB with 2\,g/l PPO was loaded at BNL with 2\% mass fraction of $^{\mathrm{nat}}$Sm which contains the $\alpha$--emitter $^{147}$Sm with a natural abundance of 14.99(18)\%. This isotope decays into $^{143}$Nd with a half--life of  \mbox{$1.06 \times10^{11}$\,y} \cite{eks04}. The kinetic energy of the $\alpha$--par\-ticles is 2.248\,MeV, which is lower than the energies accessible by \linebreak $^{12}$C($n$,$\alpha$)$^9$Be reactions. The samarium experiment is thus complementary to the previously described measurement. The fact that only one $\alpha$--par\-ticle energy is observable enhances the sensitivity to the respective \mbox{energy} compared to the situation with multiple energy peaks overlapping in the visible spectrum. To ensure that no background resulting from the radioactivity of the detector and environment fakes an $\alpha$--peak, an independent measurement with unloaded LAB + 2\,g/l PPO was conducted.

The high concentration of Sm increases the density of the full liquid scintillator cocktail to about 0.99\,g/cm$^3$. As no direct measurement of the density was done at the time, it was calculated assuming a LAB density of 0.86\,g/cm$^3$ for the LAB cocktail \cite{p500q} and a Sm density of 7.54\,g/cm$^3$ \cite{gre88}.

\subsection{Experimental setup}
To measure the $\alpha$--par\-ticle light output, the liquid scintillator was filled into a UV transparent cylindrical fused silica cuvette\footnote{Hellma Analytics.} with a length of 100\,mm and a diameter of 19\,mm. The cuvette can thus hold about 28\,ml Sm--loaded LAB. To avoid oxygen in the cuvette, the filling was performed in a nitrogen atmosphere. The LAB solution itself was de--oxygenated by bubbling it for 30\,min with gaseous nitrogen. The filled cuvette was covered with Teflon (PTFE) tape to improve the reflectivity at the cuvette walls. Each front of the cuvette was coupled to an R2059-01 PMT using UV transparent Baysilone M 200.000 silicon grease. This type of phototube is equipped with a fused silica window and provides an increased UV sensitivity.  An active vol\-tage divider guaranteeing long term stability in gain was used for the high voltage supply of the PMTs \cite{voltage_stabilisation}. The setup with two PMTs operating in coincidence mode enabled the suppression of thermal noise and improves the position dependent resolution. The cuvette was covered with aluminum foil and a heat--shrink tube, to screen it from light, and was tested to be light tight.

The signal of each PMT was split. One signal was directed to a fast Acqiris DC--282 digitizer, the other one was fed into an Ortec 584 constant fraction discriminator (CFD). The CFD threshold was set close to the electronic noise level, to allow the detection also of small pulses. The signals from the CFD were fed into a fast coincidence unit with a coincidence time window of 50\,ns. In case the coincidence condition was fulfilled, data taking was triggered and the time dependent, digitalized pulse was stored. The coincidence events were stored event--by--event. Each stored pulse was subsequently integrated, providing the $PH$ of the event in arbitrary units (arb.u.). The $PH$ scale is calibrated with standard $\gamma$--ray sources. 

\subsection{Calibration}
\label{ssec:smcal}
 \bildb{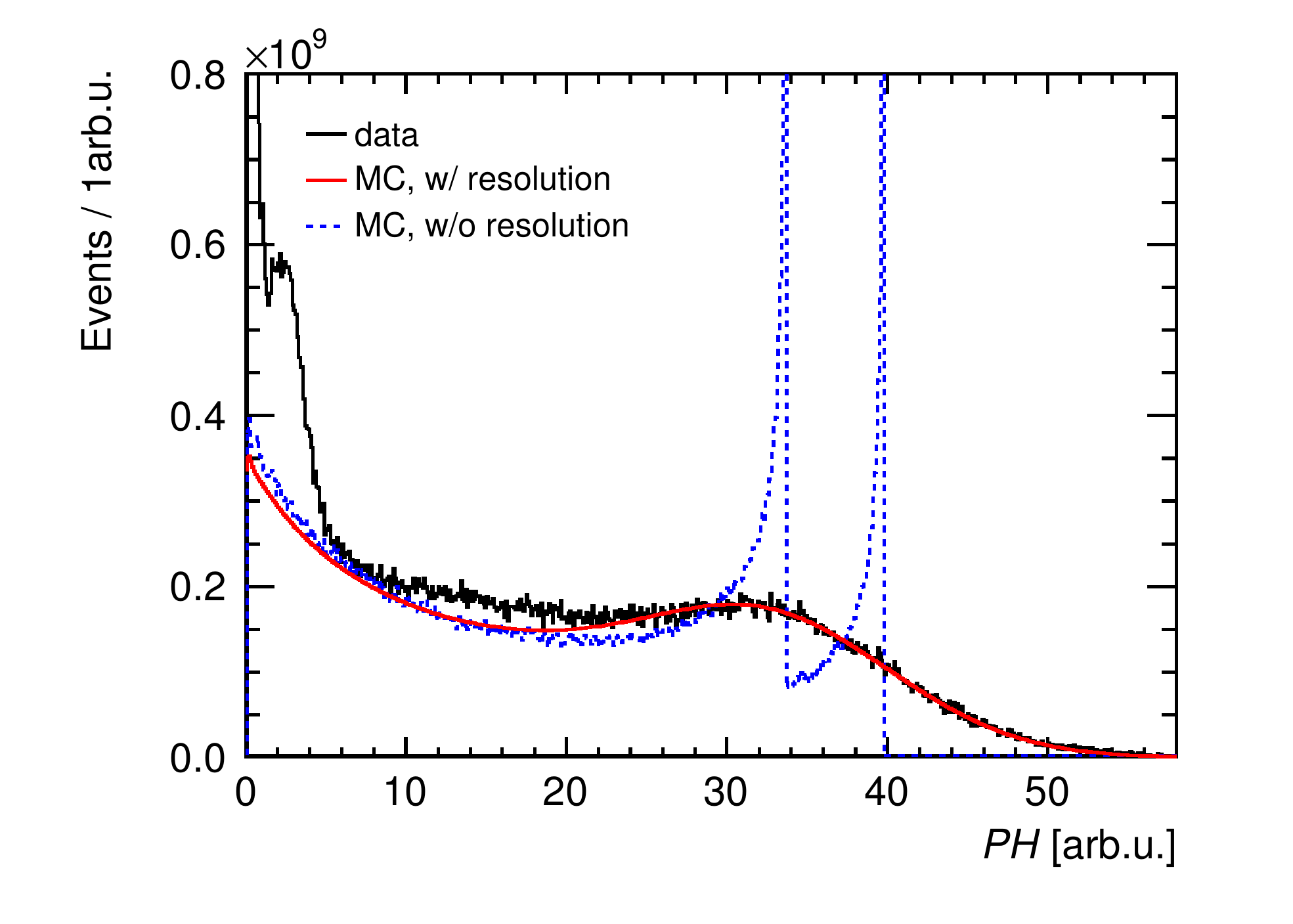}{0.5\textwidth}{Distribution of the pulse--height $PH$ measured with a $^{60}$Co calibration source. The data were taken with LAB + 2\,g/l\,PPO + 2\% \textsuperscript{nat}Sm. The peak between (1-6)\,arb.u. is due to the $\alpha$--decay of $^{147}$Sm. Also shown is the Monte Carlo (MC) $PH$ distribution before and after folding it with the detector resolution function}{0.48\textwidth}{fig:Co60}{}{h!}
Three different sources were used for calibration: \textsuperscript{137}Cs, \textsuperscript{60}Co and \textsuperscript{166m}Ho. The latter two sources provide several $\gamma$--rays, which are too close in energy to be resolved with the given setup and scintillator. The simulation cannot be fitted to the data spectrum around each Compton edge independently. This instance is demonstrated in Fig.~\ref{fig:Co60} for the two dominant $\gamma$--rays of \textsuperscript{60}Co with an energy of 1.333\,MeV and 1.173\,MeV. Note that a simplified MC simulation is used for this experiment which does not consider multiple scattering, any other interaction than Compton scattering, secondary photons or surface effects. The only detector effect con\-sidered is the $PH$ resolution. This treatment does not fully reproduce the data below the Compton edges including the distinct peak between (1-6)\,arb.u. from the $\alpha$--decay of $^{147}$Sm. However, the region of interest for the calibration is the region around the Compton edges, which is well--described by this calculation. 

As a consequence of the unresolved single Compton edges, only the weighted mean value of the $\gamma$--energies is used per source, i.e. $E_\gamma = 1.041$\,MeV for $^{60}$Co and \mbox{$E_\gamma = 0.571$\,MeV} for $^{166\mathrm{m}}$Ho. In total three calibration data points $PH_i$ are thus accessible. A fit of Eq.~\ref{equ:gcal} to these data points results in the calibration parameters \linebreak \mbox{$m=(38.1 \pm 1.8)$\,arb.u./MeV} and $a=(-2.8 \pm 1.2)$\,arb.u. using Sm--loaded LAB and \mbox{$m=(60.0 \pm 9.6)$\,arb.u./MeV} and $a=(-2.7 \pm 0.4)$\,arb.u. using unloaded LAB.

\subsection{Determination of the $^{147}$Sm $\alpha$--peak position}
The calibrated background distribution measured with unloaded LAB is shown in Fig.~\ref{fig:BGonly}. It is expected to mainly consist of events induced by $\beta$--particles and $\gamma$--rays of the natural $^{238}$U and $^{232}$Th decay chains. This natural background reveals no distinct peak structure and has a rate of $(13.76\pm0.02)$\,1/s, observed over the full dynamic measurement range. The background rate above 75\,keV, the detection threshold within the measurement with Sm--loaded LAB, amounts to $(8.91\pm0.01)$\,1/s. 

\bildb{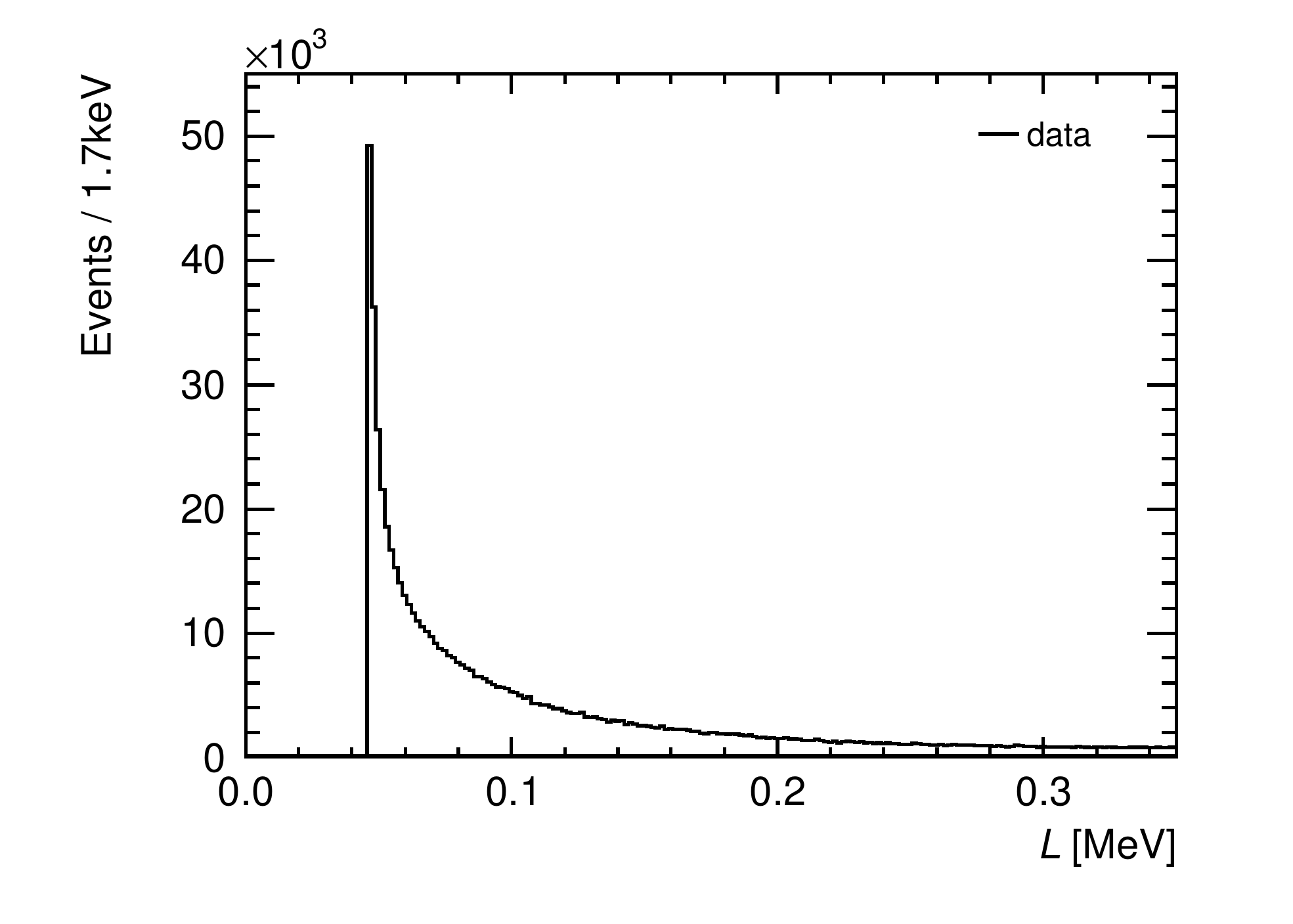}{0.5\textwidth}{Distribution of the total light yield $L$ in units of electron--equivalent energy of background events measured with LAB + 2\,g/l PPO. The background is dominated by $\beta$ and $\gamma$ events from the \mbox{natural} radioactivity of the detector material and from ambient $\gamma$--radiation}{0.48\textwidth}{fig:BGonly}{}{h!}
The light yield distribution measured with Sm--loa\-ded LAB is shown in Fig.~\ref{fig:SmQspec}. The observed event rate is $(75.80\pm0.04)$\,1/s before subtraction of the natural background and $(66.89\pm0.04)$\,1/s after subtraction of the natural background. The $^{147}$Sm activity in 28\,ml of 2\% \textsuperscript{nat}Sm--loaded LAB is about 69.7\,Bq. The observed $^{147}$Sm decay rate is lower than the $^{147}$Sm activity mainly due to a finite detection efficiency. 

Note that the bin widths in Fig.~\ref{fig:BGonly} and Fig.~\ref{fig:SmQspec} vary due to the different calibration parameters. The difference has been accounted for within the background subtraction. The incomplete background reduction in Fig.~\ref{fig:SmQspec} is expected, since the loaded LAB cocktail is expected to have a higher level of impurities from the U and Th chain. Though \textsuperscript{nat}Sm itself also contains another unstable isotope, $^{148}$Sm, with an abundance similar to the one of $^{147}$Sm, it does not sig\-ni\-fi\-cant\-ly contribute to the measurement, since its half--life of $7 \times10^{15}$\,y is nearly four orders of magnitude larger \cite{eks04}. The chemical purity\footnote{ACS grade from Sigma--Aldrich.} of the \textsuperscript{nat}Sm used is $\geq$99\% and no further rare Earth elements are expected to contribute to the background spectrum. Note that despite the residual natural background, a distinct $\alpha$--peak due to the decays of $^{147}$Sm isotopes is clearly visible. 
\bildb{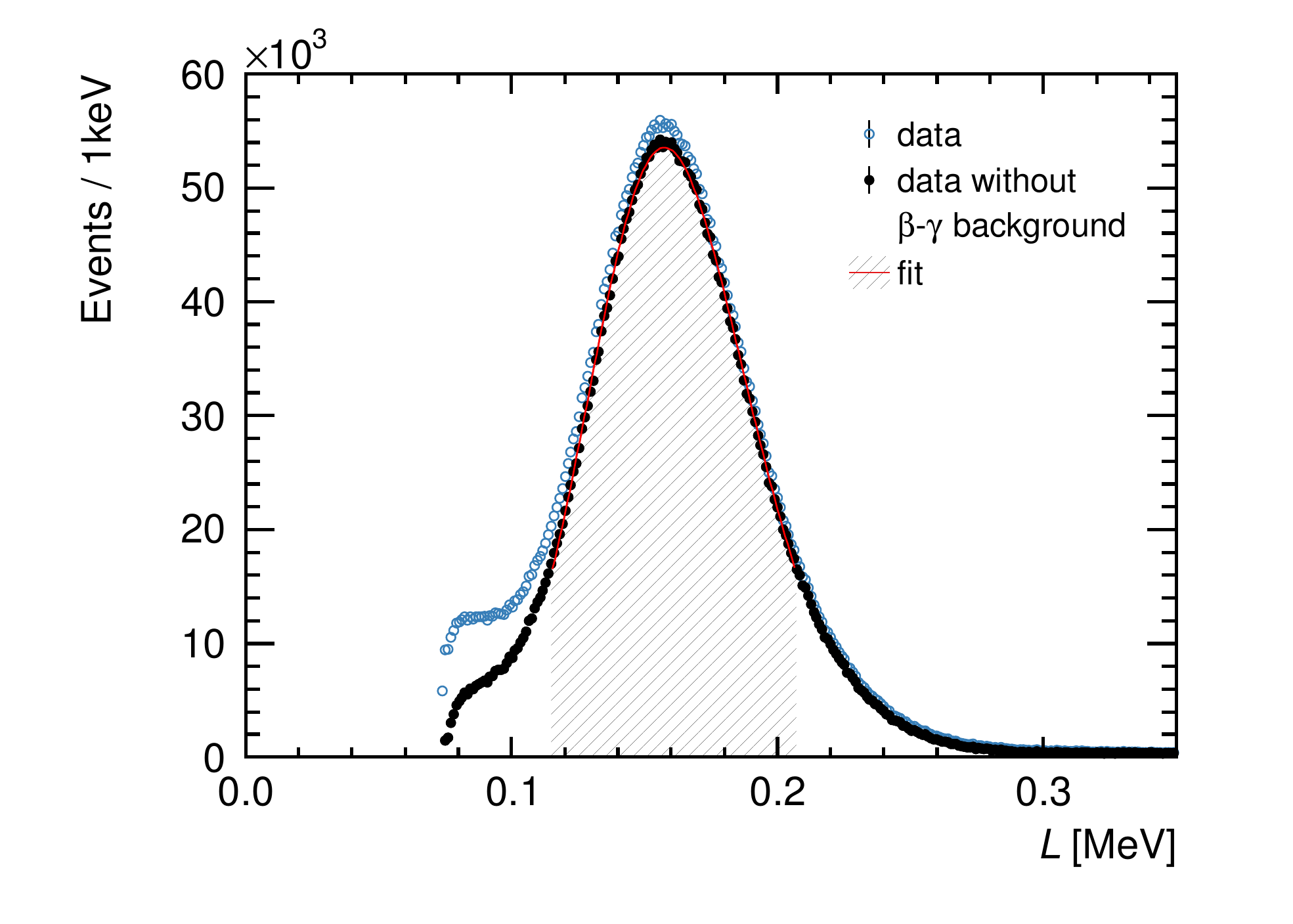}{0.5\textwidth}{Distribution of the total light yield $L$ in units of electron--equivalent energy measured with LAB + 2\,g/l PPO + 2\% \textsuperscript{nat}Sm before and after background subtraction. This background was measured with unloaded LAB. The peak results from the $\alpha$--decay of $^{147}$Sm and is fitted with an asymmetric Gaussian distribution}{0.48\textwidth}{fig:SmQspec}{}{h!}

The value of $L$ corresponding to an $\alpha$--par\-ticle \mbox{energy} of 2.248\,MeV is obtained by locating the $^{147}$Sm $\alpha$--peak in Fig.~\ref{fig:SmQspec}. Due to the residual background the peak is not fully Gaussian, but slightly asymmetric. Thus a generalization of the Gaussian distribution function,
\begin{equation}
\label{equ:skewnormal}
f(L) = \frac{e^{-\frac{(L-\xi)^2}{2\omega^2}}}{\sqrt{2\pi}\omega}\,\,\times \,\,\mathrm{erfc}\left[ -\frac{s(L-\xi)}{\sqrt{2} \omega} \right],
\end{equation}
is used, which allows for non--zero skewness \cite{arn02,azz14}. The function "erfc$(x)$" is the complementary Gauss error function, which equals one if the distribution is sym\-met\-ric. In this case, the shape parameter $s$ is zero. The parameters $\xi$ and $\omega$ approximate the mean value and standard deviation of the distribution as it approximates a normal distribution. The best fit of Eq.~\ref{equ:skewnormal} is used instead of a centroid calculation, since it revealed to be more robust and furthermore reproduces the peak region well, as shown in Fig.~\ref{fig:SmQspec}. 

The $\chi^2$ value of the best fit distribution over the number of degrees of freedom amounts to 337/83. The position of the ma\-xi\-mum of this distribution corresponds to the value of $L$ at  2.248\,MeV and is determined to be $L = (0.157\pm0.032)$\,MeV. The given total uncertainty is obtained by error propagation, considering the uncertainty on the PMT gain, the uncertainty on the calibration parameters $m$ and $a$ (see Sec.~\ref{ssec:smcal}) and the fit uncertainty. The PMT gain variations are measured to be less than $\pm 1\%$. The fit uncertainty is $1\times10^{-4}$\,MeV.
According to the presented measurement, the $\alpha$--par\-ticle light output at $E=2.248$\,MeV is quenched by a factor of $14.3\pm2.9$ compared to electrons. 
A $\chi^2$ fit of Birks' law Eq.~\ref{equ:birks} to the obtained data point yields $kB = (0.0066 \pm 0.0016)$\,cm/MeV. Also in this case, $(\mathrm{d}E/\mathrm{d}x)(E)$ is calculated using \textsc{srim}. This fit is only possible, since $L(\mathrm{0\,MeV})= 0$\,MeV, which fixes the fit at this point. The given total uncertainty on $kB$ is the quadratic sum of the individual contributions and includes a 10\% uncertainty on the calculated Sm--LAB density. This uncertainty is chosen arbitrarily large and amounts to $\pm 0.0007$\,cm/MeV. Thus, an accurate measurement of the density can reduce the total uncertainty to 0.0014\,cm/MeV. The data point and best fit function is presented and discussed in Sec.~\ref{sec:ThreeAlpha}.

\section{Measurement of $\alpha$--par\-ticle quenching using scintillator internal radioactivity and the SNO+ detector}
\label{sec:bucket}
The third experiment discussed here is the bucket source experiment. It was conducted in 2008 during the transition phase from the SNO \cite{bog00} to the SNO+ \cite{loz15} experiment. In this phase, the entire SNO/SNO+ detector was filled with water. The results are presented here by courtesy of the SNO+ collaboration.

\subsection{Experimental setup}

The SNO+ detector is installed in a barrel--shaped 34\,m deep and $\leq22$\,m wide cavity. The center of the detector is formed from a 12\,m diameter acrylic vessel (AV) of 5\,cm thickness, connected to the deck area via a cylindrical neck and supported by a hanging rope system. It is surrounded by a PMT support structure (PSUP), a 17.8\,m diameter geodesic stainless steel frame holding more than 9400 Hamamatsu 8$^{\prime\prime}$--R1408--PMTs. Each PMT is equipped with a reflective collar. This configuration yields a solid angle coverage of about 54\% \cite{bog00}. The former SNO detector was a Cherenkov detector, optimized for the detection of UV photons. 

The bucket source measurement used a cylindrical UV transparent acrylic flask, the so--called bucket, housing about 1\,l of liquid scintillator. It was deployed into the water inside the AV. The investigated scintillator is raw LAB with 2\,g/l PPO.

The number $N_\mathrm{hit}$ of PMTs firing in an event is used as a measure of $PH$. This number is corrected, accounting for the number of non--working PMTs and the number of PMTs that were hit more than once during one event. One event refers to a trigger window of 400\,ns. The $PH$ scale is calibrated using an americium--beryllium (AmBe) source.

\subsection{Calibration}
 \bildb{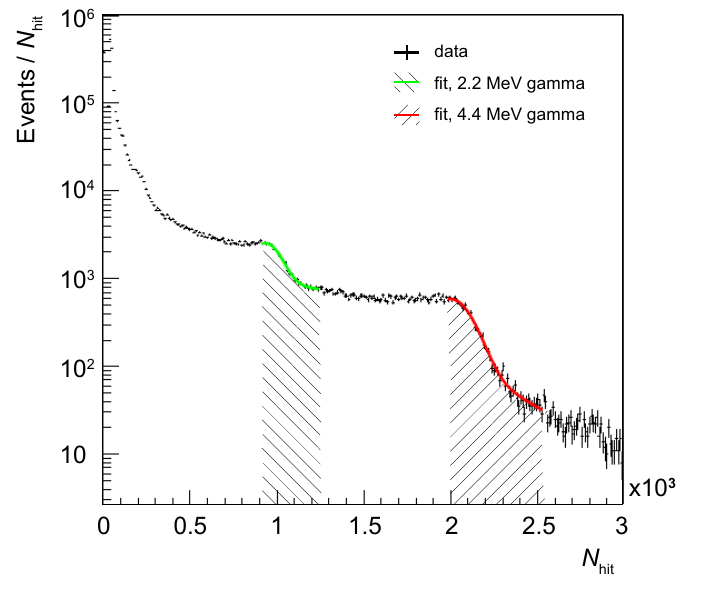}{0.5\textwidth}{$N_{\mathrm{hit}}$ distribution of an AmBe calibration source measured with LAB + 2\,g/l PPO. Also shown are fits to the Compton edges resulting from the direct 4.4\,MeV $\gamma$--rays and the delayed 2.2\,MeV $\gamma$--rays from neutron capture on $^1$H}{0.48\textwidth}{fig:AmBe}{}{h!}
To determine the light output of electrons, an AmBe source was attached to the bucket, producing neutrons and 4.4\,MeV $\gamma$--rays in $^{9}$Be($\alpha$,$n \gamma$)$^{12}$C reactions. After thermalization, neutrons are mostly captured on $^1$H, producing a deuteron and a 2.2\,MeV $\gamma$--ray. The two distinct Compton edges of known electron energy, shown in Fig.~\ref{fig:AmBe}, enable the calibration of the light output scale, using Eq.~\ref{equ:gcal} with $PH=N_{\mathrm{hit}}$. Analogous to the calibration described in Sec.~\ref{ssec:nbeamgcal}, each Compton edge is located in the $N_{\mathrm{hit}}$ spectrum by a fit including the effect of the detector resolution and by in\-spec\-ting the underlying distribution without the impact of the resolution. The resulting best fit calibration parameters of Eq.~\ref{equ:gcal} are $m = (489\pm 3)$\,/MeV and $a = -28.6\pm 8.5$.

\subsection{Determination of the $\alpha$--par\-ticle response}

Internal $\alpha$--par\-ticles were produced by dissolved $^{222}$Rn gas in the scintillator. $^{222}$Rn itself emits with a branching ratio of 99.9\% $\alpha$--par\-ticles with an energy of 5.49\,MeV, producing $^{218}$Po which also decays under the emission of an $\alpha$--par\-ticle. The $^{218}$Po $\alpha$--par\-ticle has an energy of 6.00\,MeV and the branching ratio for this decay is 99.98\%. The third kind of $\alpha$--par\-ticles identified within the bucket source data has an energy of 7.69\,MeV and results from decaying $^{214}$Po, a daughter nuclide further down the natural $^{238}$U decay chain. This decay mode is the only one of $^{214}$Po.
 \bildb{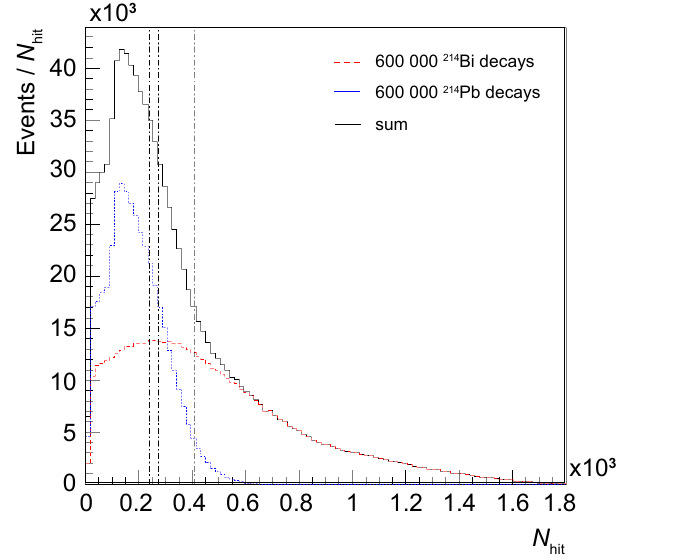}{0.5\textwidth}{Simulated $N_{\mathrm{hit}}$ distributions of backgrounds from $^{214}$Pb and $^{214}$Bi decays. The vertical dashed--dotted lines, in order of increasing $N_{\mathrm{hit}}$ value, show the approximate positions of $^{222}$Rn, $^{218}$Po and $^{214}$Po $\alpha$--peaks, respectively}{0.48\textwidth}{fig:BetaGamma}{}{h!}

$^{214}$Po has a very short half--life $T_{1/2}$ of about 164.3\,$\upmu$s. $^{214}$Po events are thus easily identified making use of their time coincidence with the $\beta$--decay of the parent isotope, $^{214}$Bi (see also \cite{loz15}). To locate the $^{218}$Po and $^{222}$Rn lines, the background, which is dominated by $\beta$--particles and $\gamma$--rays from $^{214}$Pb and $^{214}$Bi decays, is first subtracted. The shape of the corresponding background sum spectrum is determined via simulation, which was performed with the SNO software package, SNOMAN \cite{bog00}, which was kindly made available by the SNO collaboration. Figure~\ref{fig:BetaGamma} shows 600\,000 simulated $^{214}$Bi and $^{214}$Pb decays. For comparison, the approximate positions of the $^{222}$Rn, $^{218}$Po and $^{214}$Po $\alpha$--peaks are indicated by vertical lines. It is found that the background decreases smoothly in the region where the $\alpha$--peaks are expected. The shape is well--described by a 6$^{\mathrm{th}}$ order polynomial.

\begin{figure}
	\centering
	\subfloat[Before background subtraction]{\label{fig:Bucket:1}\includegraphics[width=0.5\textwidth]{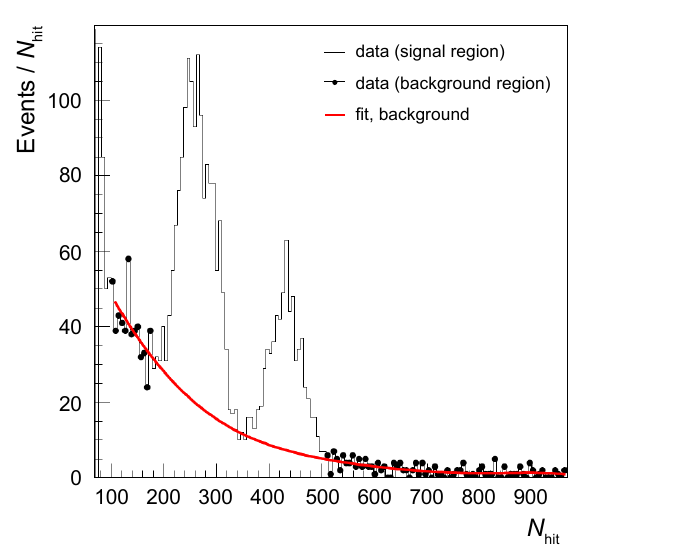}}\qquad
	\subfloat[After background subtraction]{\label{fig:Bucket:2}\includegraphics[width=0.5\textwidth]{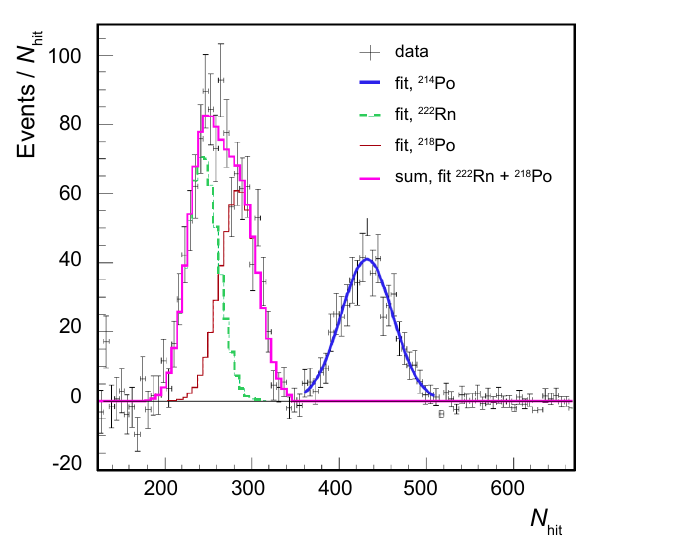}}
	\caption{\label{fig:Bucket} Exemplary $N_{\mathrm{hit}}$ distribution measured during one of the bucket source data taking runs. The fit of a $6^{\mathrm{th}}$ order polynomial to the natural background, mainly from $^{214}$Pb and $^{214}$Bi decays, is shown in (a). This background is subtracted in (b), where three Gaussian distributions are fitted to the $\alpha$--peaks from $^{222}$Rn, $^{218}$Po and $^{214}$Po}
\end{figure}

The bucket source data were taken in several runs. Figure~\ref{fig:Bucket} shows data from one run, as an example, before and after background subtraction in the region of \mbox{interest} around the $\alpha$--peaks. Furthermore, in Fig.~\ref{fig:Bucket:2} the first peak is fitted with two Gaussian components, corresponding to $^{222}$Rn and $^{218}$Po. The respective normalizations are constrained to be equal, assuming equilibrium. The means and standard deviations, instead, are allowed to float. The second peak, which is visible at a higher $N_{\mathrm{hit}}$ value, is fitted with a single Gaussian function, accounting for the $^{214}$Po line with all parameters floating freely. The best fit normalization was found to be consistent with the equilibrium assumption.

\setlength{\tabcolsep}{6pt}
\begin{table}[htbp]
\begin{center}
\caption[]{\label{tab:bucket}  Mean $N_{\mathrm{hit}}$ values measured at an $\alpha$--par\-ticle energy $E_\alpha$ and light output $L$ in units of electron--equivalent energy after calibration. The uncertainty on $L$ results from error propagation of the uncertainties on $N_{\mathrm{hit}}$ and the calibration parameters $m$ and $a$. The scintillator used is LAB + 2\,g/l PPO}
\begin{tabular*}{0.44\textwidth}{cccc}
\hline\noalign{\smallskip}
  $\alpha$--emitter & $E_\alpha$ [MeV] & $N_{\mathrm{hit}}$ & $L$ [MeV] \\\noalign{\smallskip}
\noalign{\smallskip}\hline\noalign{\smallskip}
$^{222}$Rn & 5.49 & $246.4\pm2.2$ & $0.56\pm0.02$ \\\noalign{\smallskip} 
$^{218}$Po & 6.00 & $284.2\pm2.1$ & $0.64\pm0.02$ \\\noalign{\smallskip} 
$^{214}$Po & 7.69 & $422.2\pm2.6$ & $0.92\pm0.02$ \\\noalign{\smallskip} 
\hline
\end{tabular*}
\end{center}
\end{table}

The best fit values of the Gaussian mean and the respective uncertainties provide the three data points listed in Tab.~\ref{tab:bucket}. Also listed are the values of the resulting light output $L$ in units electron--equivalent energy. The uncertainty of $L$ is obtained by propagating the uncertainties of the calibration parameters and the observed $N_{\mathrm{hit}}$ value. A fit of Birks' law Eq.~\ref{equ:birks} to the three values of $L$ yields $kB = (0.0072 \pm 0.0003)$\,cm/MeV. Also in this analysis, $(\mathrm{d}E/\mathrm{d}x)(E)$ is calculated using \textsc{srim}. The best fit function is presented together with the bucket source data points in the next section.

\section{Comparison of the results from the three independent $\alpha$--par\-ticle quenching experiments}
\label{sec:ThreeAlpha}

\begin{figure}
\centering
\subfloat[Neutron beam sample: 2\,g/l\,PPO+15\,mg/l\,bis--MSB]{\label{fig:ThreeAlpha:1}\includegraphics[width=0.5\textwidth]{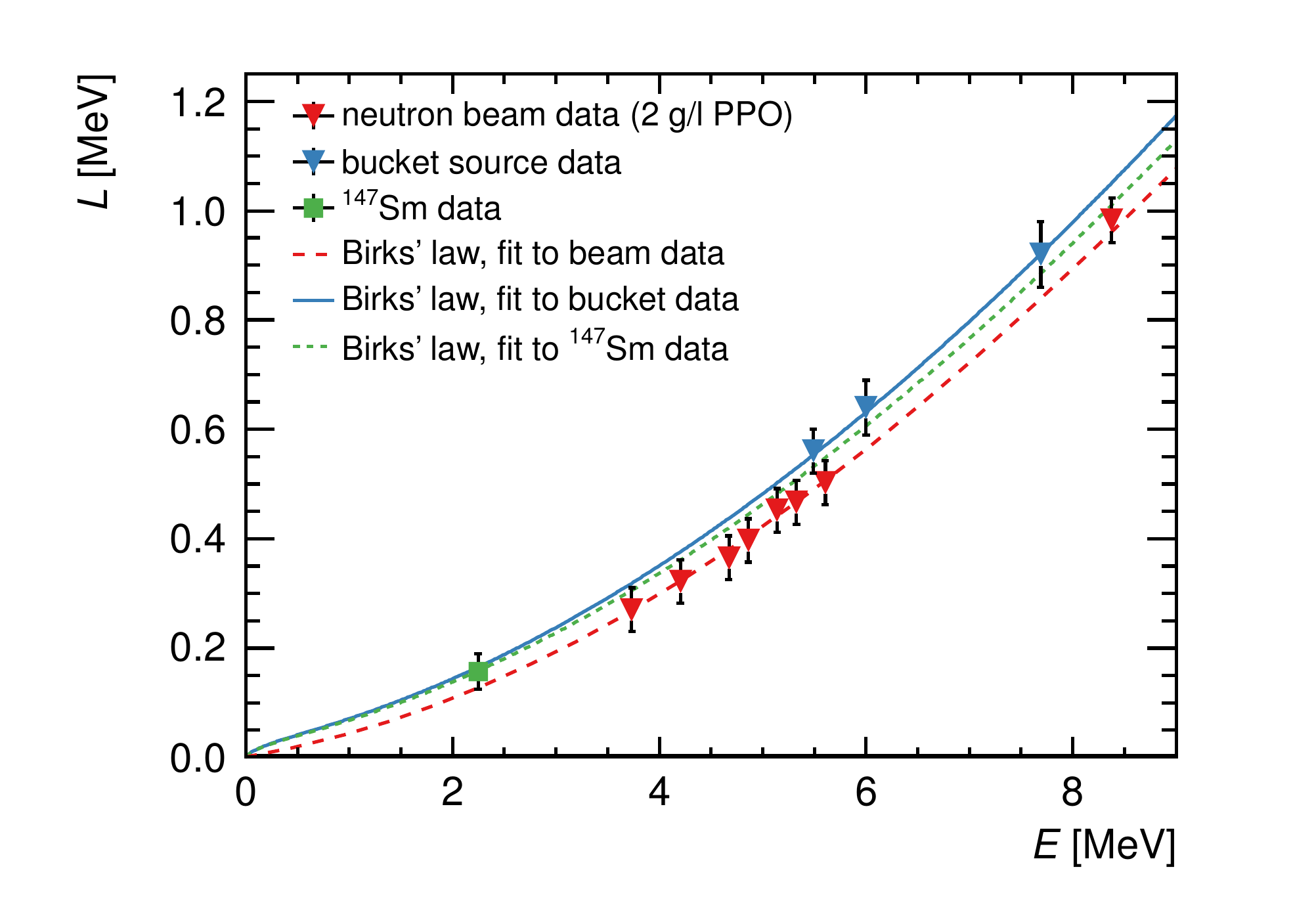}}\qquad
\subfloat[Neutron beam sample: 3\,g/l\,PPO+15\,mg/l\,bis--MSB]{\label{fig:ThreeAlpha:2}\includegraphics[width=0.5\textwidth]{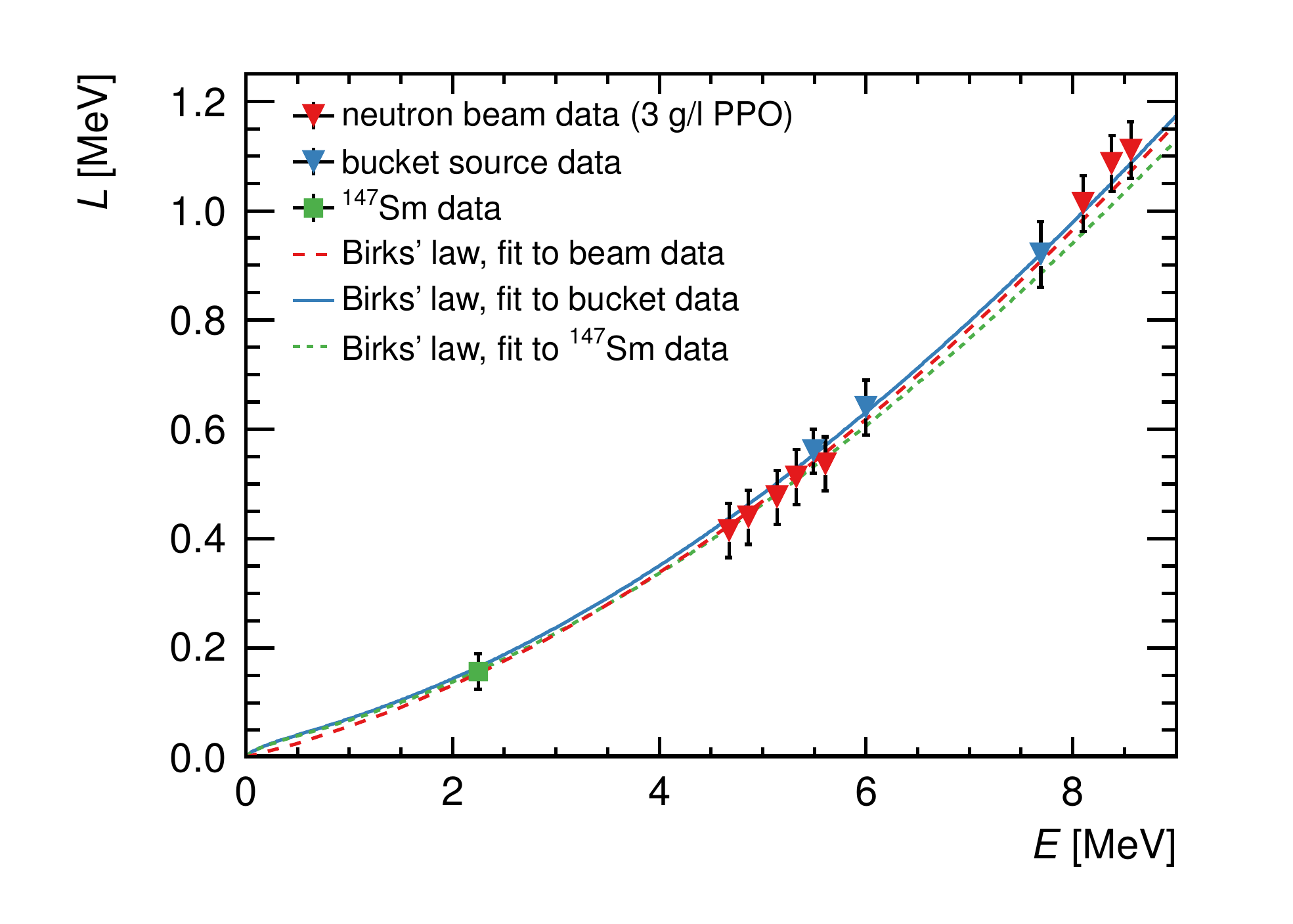}}
\caption{\label{fig:ThreeAlpha} Light output $L$ of $\alpha$--par\-ticles in units of electron--equivalent energy as a function of the kinetic energy $E$. The LAB scintillator used for the data taken with the neutron beam contained 2\,g/l PPO + 15\,mg/l bis--MSB (a) or 3\,g/l PPO + 15\,mg/l bis--MSB (b). For an easier comparison, the same data from the bucket source and the samarium ex\-peri\-ment is shown in both panels. The LAB scintillators used for these experiments contained 2\,g/l PPO as the only fluor. Also shown is a fit of Birks' law Eq.~\ref{equ:birks} to each set of data points}
\end{figure}

All light output data points, measured with the three presented experiments, are shown in Fig.~\ref{fig:ThreeAlpha}. For better visibility, the results from the two LAB samples used for the neutron beam experiment are compared to the other two experiments individually in Fig.~\ref{fig:ThreeAlpha:1} and \ref{fig:ThreeAlpha:2}. Additionally shown are the individual best fit light output functions, parameterized by Birks' law Eq.~\ref{equ:birks}. The corresponding best fit values are summarized in Tab.~\ref{tab:allaqu}. 

\setlength{\tabcolsep}{4pt}
\begin{table}[htbp]
\begin{center}
\caption[]{\label{tab:allaqu}  Quenching parameter $kB$ for $\alpha$--par\-ticles in LAB based scintillator determined in three independent experiments. Also given in the table are the total uncertainties on $kB$ as well as the $\chi^2$ values over the number of degrees of freedom(ndf). The experiments are described in the text}
\begin{tabular*}{0.48\textwidth}{ccc}
\hline\noalign{\smallskip}
Experiment & $kB$ [cm/MeV]  & $\chi^2/\mathrm{ndf}$\\
(Fluors) &  & \\
\noalign{\smallskip}\noalign{\smallskip}
\hline
neutron beam & 0.0076\,$\pm$\,0.0003 & 0.73 \\\noalign{\smallskip}
(2\,g/l\,PPO, 15\,mg/l\,bis-MSB) &  &  \\\noalign{\smallskip}\hline
neutron beam & 0.0071\,$\pm$\,0.0003 & 1.36 \\\noalign{\smallskip}
(3\,g/l\,PPO, 15\,mg/l\,bis-MSB) &  &  \\\noalign{\smallskip}\hline
bucket source & 0.0072\,$\pm$\,0.0003 & 0.02 \\\noalign{\smallskip}
(2\,g/l\,PPO) &  &  \\\noalign{\smallskip}\hline
samarium & 0.0066\,$\pm$\,0.0016 & $2\times10^{-9}$\\\noalign{\smallskip}
(2\,g/l\,PPO) &  &  \\\noalign{\smallskip}
\hline
\end{tabular*}
\end{center}
\end{table}

The best fit to the $^{147}$Sm data reveals an overly small reduced $\chi^2$ value due to the fact that only a single data point exists which additionally has a large uncertainty. Furthermore, the obtained best fit $kB$ value has the largest uncertainty and is smaller than the results from the neutron beam and bucket source experiments. The large uncertainty on $kB$ results mainly from the uncertainty on the calibration parameters. It could be reduced in a future experiment by the use of an additional fluor, like bis--MSB, to improve the light output resolution and by the use of a greater variety of calibration sources with at best well--separated Compton edges. It is important to note, though, that all best fit $kB$ values, determined in the three different experiments, agree well within their total uncertainties.

The neutron beam data, observed with LAB + \linebreak 3\,g/l\,PPO + 15\,mg/l\,bis--MSB, is in excellent agreement with the bucket source data (see Fig.~\ref{fig:ThreeAlpha:2}) and also the fitted light output functions agree remarkably well. A slight tension among data is observed in Fig.~\ref{fig:ThreeAlpha:1} though, where LAB + 2\,g/l\,PPO + 15\,mg/l\,bis--MSB was used in the neutron beam experiment. The neutron beam measurement is dominated by systematic uncertainties \cite{kro13}, where the systematic effects are mostly correlated. The correlations have been taken care of by including all sources of systematics as nuisance parameters in the $\chi^2$ fit, as discussed in \cite{kro13}. Inspecting the individual pulls of the systematics revealed that the main contribution to the $\chi^2$ is due to the prompt $\gamma$--peak centroid position in the TOF spectrum and the slight non--linearity of the time--to--amplitude converter (TAC) used for the TOF measurement. Thus the small tension among data is due to a systematic uncertainty in the TOF measurement. It should be noted, though, that the corresponding neutron beam data obtained with LAB + 2\,g/l\,PPO + 15\,mg/l\,bis--MSB still agrees with the bucket source data within their standard measurement uncertainties. Also the cor\-res\-pon\-ding best fit $kB$ values of \mbox{$(0.0076 \pm 0.0003$)\,cm/MeV} and $(0.0072 \pm 0.0003$)\,cm/MeV, and thus the resulting functions $L(E)$, do not significantly deviate from each other.

\section{Comparison of the results from the simultaneous $\alpha$--par\-ticle and proton quenching measurements}
\label{sec:AlphaProton}

The $\alpha$--par\-ticle (see Sec.~\ref{sec:nbeam}) and proton \cite{kro13} quenching measurements using a neutron beam provide ideal conditions to test the hypothesis of a unique quenching pa\-ra\-me\-ter $kB$ for different ions in the same liquid scintillator. This hypothesis is subject to most present investigations \cite{tre13} and important for all liquid scintillator particle detectors. The $\alpha$--par\-ticle quenching data was taken with the same detector and data acquisition system and was analyzed using the same method and assumptions as the proton quenching data discussed in \cite{kro13}. Furthermore, the data was taken simultaneously and thus with the same scintillator filling. Hence, neither temperature changes, nor a different amount of impurities like oxygen, nor different aging could explain differences observed for dif\-fe\-rent particle types. Finally, protons and $\alpha$--par\-ticles are both created inside the scintillator volume, thus avoiding surface effects. 

The results for both particle types and scintillator samples are compared in Tab.~\ref{tab:apqu}. While the presented results all agree for the same particle type measured with dif\-fe\-rent scintillators based on the same solvent, the $kB$ values for protons\footnote{It should be noted that \cite{kro13} considers a quadratic correction term within Birks' law, parameterized by $C$. $C$ was found to be consistent with zero. It was verified for the present article that neglecting the correction term does not change the best fit $kB$ values.} and $\alpha$--par\-ticles in the same sample deviate by more than $5\sigma$. The light output of the two particle types cannot be described by the same Birks parameter in the presented neutron beam measurements.

\setlength{\tabcolsep}{26pt}
\begin{table}[htbp]
\begin{center}
\caption[]{\label{tab:apqu}  Quenching parameter $kB$ for protons and $\alpha$--par\-ticles in LAB based scintillator resulting from simultaneous neutron beam measurements. Also given are the total uncertainties. The proton quenching results are taken from \cite{kro13}}
\begin{tabular*}{0.42\textwidth}{cc}
\hline\noalign{\smallskip}
Particle & $kB$ [cm/MeV]  \\
\noalign{\smallskip}\noalign{\smallskip}
\hline
\multicolumn{2}{c}{LAB + 2\,g/l PPO + 15\,mg/l bis--MSB} \\
\hline\noalign{\smallskip}
Proton & 0.0097\,$\pm$\,0.0002 \\\noalign{\smallskip} 
$\alpha$--par\-ticle & 0.0076\,$\pm$\,0.0003 \\\noalign{\smallskip}
\hline
\multicolumn{2}{c}{LAB + 3\,g/l PPO + 15\,mg/l bis--MSB} \\
\hline\noalign{\smallskip}
Proton & 0.0098\,$\pm$\,0.0003 \\\noalign{\smallskip} 
$\alpha$--par\-ticle & 0.0071\,$\pm$\,0.0003 \\\noalign{\smallskip} 
\hline
\end{tabular*}
\end{center}
\end{table}

\section{Summary and outlook}
\label{sec:summary}

This article presents the results of $\alpha$--par\-ticle quen\-ching measurements obtained with three independent experiments using LAB based liquid scintillators: the neutron beam experiment, the samarium experiment and the bucket source experiment. These experiments all use small liquid scintillator volumes, but apply different techniques. All three detectors have an increased sensitivity to UV light and thus Cherenkov light, which is ad\-di\-tio\-nal\-ly emitted when fast electrons traverse the scintillator. Nonetheless the wavelength sensitivities of the experiments are not identical. The $\alpha$--par\-ticles are internal in all presented measurements. The data analyses were performed under the same assumptions.

The energy dependent $\alpha$--par\-ticle light output is ana\-lytically described by Birks' law, which is pa\-ra\-me\-terized by $kB$. The best fit value of $kB$ is determined by independent fits to two sets of neutron beam data points, to the bucket source data points and to the single $^{147}$Sm data point. The four re\-sul\-ting $kB$ values range from $(0.0076\pm0.0003)$\,cm/MeV to $(0.0066\pm0.0016)$\,cm/MeV. All presented measurements agree and the observed $\alpha$--par\-ticle responses can be described by the same $kB$ value within the uncertainties. The widest range of $\alpha$--par\-ticle energies is covered by the neutron beam experiment. It is thus recommended to use the parameter $kB$ resulting from this experiment. The value $(0.0076 \pm 0.0003$)\,cm/MeV, observed with LAB + 2\,g/l\,PPO + 15\,mg/l\,bis--MSB, is slightly preferred over the one obtained with the LAB sample with 3\,g/l PPO + 15\,mg/l bis--MSB. The determined reduced $\chi^2$ value in the first case is with 0.73 slightly lower than the one in the second case with 1.36.

Within the neutron beam experiment, the proton light output was measured simultaneously with the $\alpha$--par\-ticle light output. The proton quenching analysis and results were previously presented in \cite{kro13}. The $kB$ value obtained for protons is $(0.0097\pm0.0002)$\,cm/MeV for LAB + 2\,g/l PPO + 15\,mg/l bis--MSB and $(0.0098\pm0.0003)$\,cm/MeV for LAB + 3\,g/l PPO + 15\,mg/l bis--MSB. These proton $kB$ values deviate from the corresponding $\alpha$--par\-ticle $kB$ values by $6.4\,\sigma$ and $5.8\,\sigma$, respectively. Hence, the proton and the $\alpha$--par\-ticle light responses cannot be reproduced with the same $kB$ value. This observation is fundamental for current and future large--scale liquid scintillator detectors, like SNO+, which require a precise modeling of the signal and background light yield distribution over several MeV of elec\-tron--equivalent energy, induced by different kinds of particles. A reliable background model is crucial for e.g. the development of most efficient background rejection techniques. 

This article demonstrates that the results obtained with the neutron beam experiment are transferable to the SNO+ experiment, despite non--identical wavelength sensitivities. This finding is very important, since it enables the transfer of also the proton quenching results \cite{kro13} to the SNO+ MC model. It furthermore allows to examine further LAB based scintillators {\it ex situ}, which are of interest for SNO+, like Te--loaded LAB \cite{loz15}.

In future laboratory experiments, the influence of the wavelength sensitivity of the detector, particularly in the UV light region, should be systematically examined. This dependence is of interest for all liquid scintillator ex\-peri\-ments in general. Besides this, the observed difference in the proton and $\alpha$--par\-ticle quenching parameters motivates systematic measurements using heavier ions under identical measurement conditions in order to investigate the dependence of $kB$ on ion properties like the mass. Though this investigation is of limited practical relevance for typical liquid scintillator experiments, it is of fundamental interest for the theory of liquid scintillation and the dependence of $kB$ on the ion type. This kind of measurement is only possible at accelerator facilities, where ion beams of sufficiently high energies can be produced to induce an observable amount of scintillation light.

\begin{small}
\begin{acknowledgements}
We acknowledge provision of the bucket source data by the SNO+ collaboration and software tools by the SNO collaboration. We further thank the mechanical workshop of the TU~Dresden for the production of the scintillator cell as well as Kai Tittelmeier, the accelerator staff of the PTB and Andreas\,Hartmann from HZDR for their support. We give thanks to Arnd S{\"o}rensen for confirming the R2059-01 PMT gain stability. The LAB solvent was provided by Petresa~Canada~Inc., B\'{e}can\-cour QC. This work has been in part supported by the Deut\-sche For\-schungs\-ge\-mein\-schaft (DFG), Germany (grant No. ZU-123/5), in part by the Science and Technology Facilities Council (STFC) of the United Kingdom (Grants no. ST/J001007/1, ST/K001329/1 and  ST/M00001X/1) and in part by the U.S. Department of Energy under Contract No. DE-SC0012704.
\end{acknowledgements}
\end{small}

\bibliographystyle{spphys} 
\bibliography{literature}

\end{document}